\documentclass[pra,amsmath,amssymb,reprint,superscriptaddress]{revtex4-1}

\usepackage{graphicx}% Include figure files
\usepackage{dcolumn}% Align table columns on decimal point
\usepackage{bm}% bold math
\usepackage[utf8]{inputenc}
\usepackage[T1]{fontenc}
\usepackage{etoolbox}
\usepackage{color}

\usepackage{txfonts}

\newcommand{\p}{\mathrm{p}}
\renewcommand{\H}{\mathrm{H}}
\newcommand{\V}{\mathrm{V}}
\newcommand{\sinc}{\mathrm{sinc}}
\newcommand{\e}{\mathrm{e}}
\newcommand{\vac}{\text{vac}}
\newcommand{\twin}{\text{twin}}
\newcommand{\bmu}{\bm{\mu}}
\newcommand{\be}{\mathbf{e}}
\newcommand{\bE}{\mathbf{E}}
\newcommand{\SE}{\mathrm{SE}}
\newcommand{\GSB}{\mathrm{GSB}}
\newcommand{\ESA}{\mathrm{ESA}}
\newcommand{\rephasing}{\mathrm{r}}
\newcommand{\nonrephasing}{\mathrm{nr}}
\newcommand{\calE}{\mathcal{E}}
\newcommand{\dt}{\Delta t}
\newcommand{\adagger}{\hat{a}^\dagger}
\renewcommand{\a}{\hat{a}}
\renewcommand{\Im}{\mathrm{Im}}
\renewcommand{\Re}{\mathrm{Re}}
\newcommand{\tr}{\mathrm{tr}}

\newcommand{\E}{\hat{E}}
\newcommand{\DQC}{\mathrm{DQC}}

\makeatletter
\def\@email#1#2{%
 \endgroup
 \patchcmd{\titleblock@produce}
  {\frontmatter@RRAPformat}
  {\frontmatter@RRAPformat{\produce@RRAP{*#1\href{mailto:#2}{#2}}}\frontmatter@RRAPformat}
  {}{}
}%
\makeatother
\begin{document}

\title{Probing exciton dynamics with spectral selectivity through the use of quantum entangled photons}
% Force line breaks with \\
\author{Yuta Fujihashi}
\affiliation{Department of Molecular Engineering, Kyoto University, Kyoto 615-8510, Japan}
\affiliation{PRESTO, Japan Science and Technology Agency, Kawaguchi 332-0012, Japan}
\email{fujihashi@uec.ac.jp}
 \altaffiliation[Present address: ]{Department of Engineering Science, The University of Electro-Communications, Chofu 182-8585, Japan}
\author{Kuniyuki Miwa}
 \affiliation{Institute for Molecular Science, National Institutes of Natural Sciences, Okazaki 444-8585, Japan}
 \affiliation{Graduate Institute for Advanced Studies, SOKENDAI, Okazaki 444-8585, Japan}
\author{Masahiro Higashi}
\affiliation{Department of Molecular Engineering, Kyoto University, Kyoto 615-8510, Japan}
\affiliation{PRESTO, Japan Science and Technology Agency, Kawaguchi 332-0012, Japan}
\author{Akihito Ishizaki}
\email{ishizaki@ims.ac.jp}
 \affiliation{Institute for Molecular Science, National Institutes of Natural Sciences, Okazaki 444-8585, Japan}
 \affiliation{Graduate Institute for Advanced Studies, SOKENDAI, Okazaki 444-8585, Japan}

\begin{abstract}
Quantum light is increasingly recognized as a promising resource for developing optical measurement techniques. Particular attention has been paid to enhancing the precision of the measurements beyond classical techniques by using nonclassical correlations between quantum entangled photons. Recent advances in quantum optics technology have made it possible to manipulate the spectral and temporal properties of entangled photons, and the photon correlations can facilitate the extraction of matter information with relatively simple optical systems compared to conventional schemes. In these respects, the applications of entangled photons to time-resolved spectroscopy can open new avenues for unambiguously extracting information on dynamical processes in complex molecular and materials systems. Here, we propose time-resolved spectroscopy in which specific signal contributions are selectively enhanced by harnessing the nonclassical correlations of entangled photons. The entanglement time characterizes the mutual delay between an entangled twin and determines the spectral distribution of the photon correlations. The entanglement time plays a dual role as the knob for controlling the accessible time region of dynamical processes and the degrees of spectral selectivity. In this sense, the role of the entanglement time is substantially equivalent to the temporal width of the classical laser pulse. The results demonstrate that the application of quantum entangled photons to time-resolved spectroscopy leads to monitoring dynamical processes in complex molecular and materials systems by selectively extracting desired signal contributions from congested spectra. We anticipate that more elaborately engineered photon states would broaden the availability of quantum light spectroscopy.
\end{abstract}

\maketitle

%\begin{quotation}
%The ``lead paragraph'' is encapsulated with the \LaTeX\ 
%\verb+quotation+ environment and is formatted as a single paragraph before the %(The \verb+quotation+ environment reverts to its usual meaning after the first sectioning command.) 
%Note that numbered references are allowed in the lead paragraph.
%
%The lead paragraph will only be found in an article being prepared for the journal \textit{Chaos}.
%\end{quotation}

%\section{\label{sec:level1}First-level heading:\protect\\ The line
%break was forced \lowercase{via} \textbackslash\textbackslash}

\section{\label{sec:level1}Introduction}

In recent years, quantum light has been recognized as an important resource for the development of quantum metrology, where nonclassical features of light are exploited to enhance the precision and resolution of optical measurements beyond classical techniques \cite{Pirandola:2018ad,Morea:2019im}.
One of the striking features of quantum light is quantum entanglement.
It is a phenomenon where the state of an entire system cannot be described as the product of the quantum states of its individual constituent particles.
For instance, the use of photon entanglement has enabled ghost imaging \cite{Pittman:1995un}, quantum imaging with undetected photons \cite{Lemos:2014qu}, quantum lithography \cite{Boto:2000qu}, cancellation of even-order dispersion \cite{Franson:1992no}, quantum optical coherence tomography \cite{Abouraddy:2002qu,Nasr:2003de,Okano:2015re}, and realization of sub-shot-noise microscopy \cite{Ono:2013ei,Triginer:2020qu,Casacio:2021qu}.

With recent advances in quantum optical technologies, entangled photons have become a promising avenue for the development of new spectroscopic techniques \cite{Gea1989:tw, Javanainen1990:li, Saleh:1998vl, Oka:2010if, Schlawin2013two, deJLeonMontiel:2019jt, Fujihashi:2020ep, Debnath:2020en, Szoke:2020fc, Mukamel:2020ej,Munoz:2021qu, Raymer:2021en, Chen:2021vi, Dorfman:2021ho, Asban:2021in, Asban:2021di, Asban:2022no, Chen:2022en, Albarelli:2023fu, Li:2023si}. It was experimentally demonstrated that nonclassical correlations between entangled photons had several advantages in spectroscopy, including sub-shot-noise absorption spectroscopy \cite{Tapster:1991su,Brida:2010exl,Matsuzaki:2022su} and increased two-photon absorption signal intensity \cite{Georgiades:1995dd, Dayan:2004kg, Lee:2006id, Upton:2013is, Varnavski:2020tw}. 
Entanglement-induced two-photon transparency \cite{Fei:1997es} and suppression of exciton transport \cite{Schlawin:2013dq} controlling the entanglement time, which is the hallmark of the non-classical photon correlation, were also theoretically investigated. 
In addition to the above advantages, the nonclassical correlations can be used to obtain spectroscopic signals with simpler optical systems compared with conventional methods \cite{Yabushita:2004hy, Kalashnikov:2016cl, Mukai:2021qu, Arahata:2022wa, Kalashnikov:2017hx, Eshun:2021in}. For example, infrared spectroscopy with visible-light source and detector was performed by exploiting entangled visible and infrared photons generated via parametric down-conversion (PDC) \cite{Kalashnikov:2016cl, Mukai:2021qu, Arahata:2022wa}. Furthermore, the Hong--Ou--Mandel interferometer with entangled photons allows the measurement of the dephasing time of molecules at the femtosecond time scale without the need for ultrashort laser pulses \cite{Kalashnikov:2017hx, Eshun:2021in}. 
Inspired by the capabilities of such nonclassical photon correlations, the applications to time-resolved spectroscopic measurements have been theoretically discussed \cite{Dorfman:2014bn, Schlawin:2016er, Zhang:2022en, Fan:2023en}. The development of time-resolved spectroscopy that enhances the precision and resolution beyond classical techniques may lead to a better understanding of the mechanism of dynamical processes in complex molecules, such as photosynthetic light-harvesting systems. In contrast to many experimental and theoretical studies on entangled two-photon absorption, only a few theoretical studies have reported the application of entangled photons to time-resolved spectroscopic measurements. There is no comprehensive understanding of which nonclassical states of light are suitable for implementing real-time observation of dynamical processes in condensed phases and which nonclassical photon correlations allow the manipulation of nonlinear signals in a way that cannot be achieved with classical pulses. 

In a previous study, we developed a theory of frequency-dispersed transmission measurement using entangled photon pairs generated via PDC pumped with a monochromatic laser \cite{Ishizaki:2020jl}. Especially, it was demonstrated that this measurement scheme enabled time-resolved spectroscopy based on monochromatic pumping when the entanglement time is sufficiently short. Chen {\it et al.} demonstrated that a similar scheme could be applied to monitor ultrafast electronic-nuclear motion at conical intersection \cite{Chen:2022mo, Gu:2023ph}. 
Moreover, the simple model calculations in Refs.~\citenum{Ishizaki:2020jl} and \citenum{Fujihashi:2021ac} suggested that for a finite value of entanglement time, the spectral distribution of the phase-matching function works as a sinc filter in signal processing \cite{Owen:2007pr}, which can be used to selectively resolve a specific region of the spectra. Therefore, this spectral filtering mechanism is expected to simplify the interpretation of the spectra in complex molecules. 

In this study, we theoretically propose a time-resolved spectroscopy scheme that selectively enhances specific signal contributions by harnessing the nonclassical correlations between entangled photons. We apply our spectroscopic scheme to a photosynthetic pigment-protein complex, and demonstrated that the phase-matching functions of the PDC in nonlinear crystals, such as periodically poled $\mathrm{KTiOPO_4}$ crystal and $\beta$-$\mathrm{BaB_2O_4}$ crystal, allow one to separately measure specific peaks of spectra by tuning the entanglement time and the central frequencies of the entangled photons. Furthermore, we investigated whether the spectral filtering mechanism could be implemented in the range of currently available entangled photon sources.

%%%%%%%%%%%
\begin{figure*}
    \centering
    \includegraphics{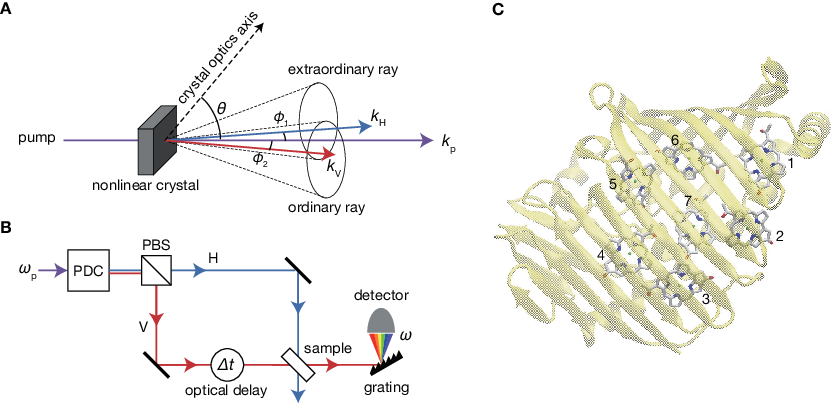}
    \caption{
	(A) Type-II PDC in a uniaxial birefringent crystal.
	In this process, one of the twin beams is created along the ordinary axis of the crystal and the other along the extraordinary axis.
	The polarizations of the two beams are orthogonal to each other.
	Here, H stands for the horizontal polarization of the extraordinary ray, and V stands for the vertical polarization of the ordinary ray.
	By selecting the angle $\theta$ of the pump propagation direction with respect to the optical axis of the crystal, one can tune the central frequencies of the entangled photon pair.
	(B) Schematic of the frequency-dispersed transmission measurement with entangled photon pairs generated via the type-II PDC pumped with a monochromatic laser of frequency $\omega_\p$ in the collinear configuration ($\phi_1=\phi_2=0$).
	The twin photons are split with a polarized beam splitter.
	The horizontally and vertically polarized photons act as the pump and probe field for the molecules, respectively.
	The probe field transmitted through the sample is frequency-dispersed, and the change in the transmitted photon number is registered as a function of frequency $\omega$ and the external delay $\dt$.
	(C) A monomer subunit of the FMO complex from {\it Chlorobium tepidum} with seven BChl{\it a} molecules. The pigments are numbered as in PDB file 1M50 \cite{CamaraArtigas:2003ch}.
    }
    \label{fig:1}
\end{figure*}
%%%%%%%%%%%%

%%%%%%%%%%%
\begin{figure}
    \centering
    \includegraphics{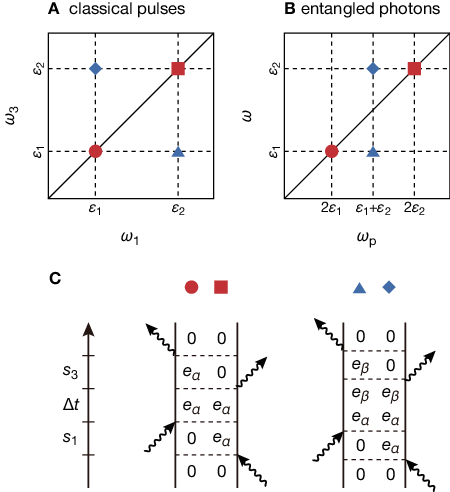}
    \caption{
    Correspondence between the classical Fourier-transformed photon-echo signal and the transmission signal with the entangled photons.
    (A) Illustration of the 2D spectrum, $\mathcal{S}_{\rm 2D}(\omega_3, \dt, \omega_1)$, of a dimer model obtained by the Fourier-transformed photon-echo measurement with the classical pulsed laser. 
    (B) The 2D spectrum, $S(\omega, \dt; \omega_\p)$, of a dimer model obtained by the transmission measurement with the entangled photons.
    In each panel, the red and blue symbols denote the signal originated from the rephasing SE pathway without and with the excitation relaxation process $e_\alpha \to e_\beta$ during $\dt$, respectively.
    For simplicity, only the peak positions of the rephasing SE signal are illustrated in panels~(A) and (B).
    (C) The corresponding double-sided Feynman diagrams.
	}
    \label{fig:2}
\end{figure}
%%%%%%%%%%%%

%%%%%%%%%%%
\begin{figure}
    \centering
    \includegraphics{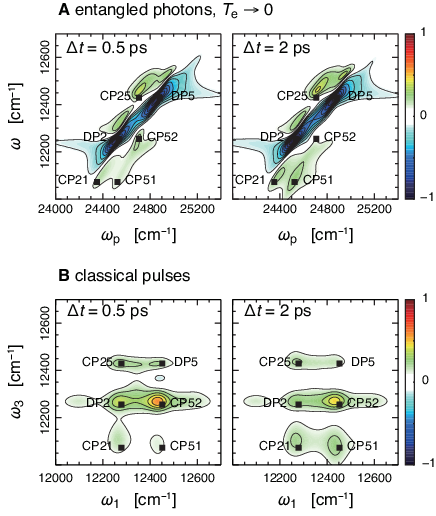}
    \caption{
	(A) Difference spectra, $\Delta S(\omega,\dt;\omega_\p) = S(\omega,\dt; \omega_\p) - S(\omega,\dt=0; \omega_\p)$, of the FMO complex with entangled photon pairs in the limit of $T_\e\to 0$.
	The waiting times are $\dt = 0.5\,{\rm ps}$ and $2\,{\rm ps}$.
	The temperature is set to $T=77\,{\rm K}$.
	The normalization of the contour plot is such that the maximum value of the spectrum at $\dt = 2\,{\rm ps}$ is unity, and equally spaced contour levels ($\pm 0.1$, $\pm 0.2$, \dots) are drawn.
	(B) Time evolution of absorptive 2D spectra, $\mathcal{S}_{\rm 2D}(\omega_3, \dt, \omega_1)$, of the FMO complex obtained with the Fourier-transformed photon-echo measurement in the impulsive limit.
	We chose the HHVV sequence as the polarization sequence of the four laser pulses.
	The normalization of the contour plot is such that the maximum value of the spectrum at $\dt = 0\,{\rm ps}$ is unity, and equally spaced contour levels ($\pm 0.1$, $\pm 0.2$, \dots) are drawn.
	}
    \label{fig:3}
\end{figure}
%%%%%%%%%%%%

%%%%%%%%%%%%%%%%%%%%%%%%%
\section{Theory}

According to the phase matching conditions, the PDC process can be triggered in different geometries: One distinguishes type-I and type-II, and type-0 down-conversion.
For simplicity, we consider a type-II PDC in a birefringent crystal \cite{mandel1995optical} because the wave vector mismatch can be well approximated to linear order in frequency \cite{Grice:1997ht,Keller:1997hj}, as described in Eq.~\eqref{eq: phase-matching-func}.
The intensity and normalized spectral envelope of the pump laser are denoted as $I_\p$ and $E_\p(\omega)$, respectively. A photon of frequency $\omega_\p$ in the pump laser is split into a pair of entangled photons whose frequencies $\omega_\H$ and $\omega_\V$ must satisfy $\omega_\p = \omega_\H + \omega_\V$
because of energy conservation. The polarizations of the generated twins are orthogonal to each other and characterized by horizontal (H) and vertical (V) polarizations. In the weak-down conversion regime, the quantum state of the twin is expressed as \cite{Grice:1997ht,Keller:1997hj}
\begin{align}
	\lvert \psi_\twin \rangle
	=
	\int d\omega_\H \int d\omega_\V
	f(\omega_\H,\omega_\V)
	\adagger_\H(\omega_\H) \adagger_\V(\omega_\V)
	\vert \vac \rangle,
	\label{eq: photon-state}
\end{align}
where the operator $\hat{a}_\lambda^\dagger(\omega)$ creates a photon of frequency $\omega$ and polarization $\lambda$ and the function $f(\omega_\H,\omega_\V)$ is the two-photon amplitude. 
For simplicity, in Eq.~\eqref{eq: photon-state}, we neglected the spatial variation of the two-photon amplitude and the spatial propagation direction is selected by the collinear configuration \cite{Schlawin:2018ci}.
In the following, we consider the electric fields inside a one-dimensional (1D) nonlinear crystal of length $L$. Thus, the two-photon amplitude is given by $f(\omega_\H,\omega_\V) = \zeta E_\p(\omega_\H+\omega_\V) \sinc[\Delta k(\omega_\H,\omega_\V)L/2]$, where $E_\p(\omega_\H+\omega_\V)$ is the normalized pump envelope and the sinc function originates from phase-matching \cite{Boyd:2003no,Graffitti:2018de}. 
Note that here $f(\omega_\H,\omega_\V)$ is not necessarily normalized, as $\int d\omega_\H \int d\omega_\V|f(\omega_\H,\omega_\V)|^2=\zeta^2$.
Expanding the wave vector mismatch $\Delta k(\omega_\H,\omega_\V)$ to first order in the frequencies $\omega_\H$ and $\omega_\V$ around the center frequencies of the generated beams, $\bar\omega_\H$ and $\bar\omega_\V$, we obtain $\Delta k(\omega_\H,\omega_\V) = (\omega_\H - \bar\omega_\H)T_\H + (\omega_\V - \bar\omega_\V)T_\V$ with $T_\lambda=(v_\p^{-1} - v_\lambda^{-1})L$, where $v_\p$ and $v_\lambda$ are the group velocities of the pump laser and one of the generated beams at central frequency $\bar\omega_\lambda$, respectively \cite{Keller:1997hj,Rubin:1994ed}. 
The central frequencies and group velocities are evaluated using the Sellmeier equations \cite{Dmitriev:2013ha}, which provide empirical relations between the refractive indices of the crystals and the frequencies of the generated beams. 
In this study, we address the case of monochromatic pumping $E_\p(\omega_\H+\omega_\V)=\delta(\omega_\H+\omega_\V-\omega_\p)$. Thus, the two-photon amplitude is recast as 
\begin{gather}
	f(\omega_\H,\omega_\V)
	=
	\zeta 
	\delta(\omega_\H + \omega_\V - \omega_\p) 
	\Phi(\omega_\V - \bar\omega_\V),
	\label{eq:two-photon-amp}
\\
	\Phi(\omega)
	=
	\sinc\frac{\omega T_\e}{2}.
	\label{eq: phase-matching-func}
\end{gather}
The so-called entanglement time $T_\e= 
\lvert
T_\H-T_\V 
\rvert
$ is the maximum time difference between twin photons leaving the crystal \cite{Saleh:1998vl}. The positive frequency component of the field operator is given by:
$	
	\hat{\bE}_\lambda^+(t)
	=
	\be_\lambda
	\int^\infty_0 d\omega\, 
	\calE(\omega)
	\a_\lambda(\omega)
	e^{-i\omega t},
$
where $\calE(\omega) \propto i\sqrt{\omega}$ and the negative frequency component is $\hat{\bE}_\lambda^-(t) = [\hat{\bE}_\lambda^+(t)]^\dagger$.
The unit vectors $\be_\H$ and $\be_\V$ indicate the directions of the horizontal and vertical polarizations, respectively. We adopt the slowly varying envelope approximation, in which the bandwidth of the field is assumed to be negligibly narrow in comparison with the central frequency \cite{mandel1995optical}. This approximation allows treating the factor $\calE(\omega)$ as a constant $\calE(\omega)\simeq \calE(\bar\omega_\lambda)$. All other constants are merged into a factor $\zeta \propto I_\p^{1/2} L \calE(\bar\omega_\H) \calE(\bar\omega_\V)$, which is regarded as the conversion efficiency of the PDC.

We consider the setup shown in Fig.~\ref{fig:1}. Twin photons were split using a polarized beam splitter. Although the relative delay between horizontally and vertically polarized photons is innately determined by the entanglement time, the delay interval is further controlled by adjusting the path difference between the beams \cite{Hong:1987gm,Franson:1989go}. This controllable delay is denoted by $\dt$ in this study. 
Direct observation of time-frequency duality of biphotons over a delay time of at least a few picoseconds has been experimentally demonstrated \cite{Jin:2018ti,Maclean:2018ul}.
The field operator is expressed as $\hat{\bE}(t) = \hat{\bE}_\H(t+\dt) + \hat{\bE}_\V(t)$, indicating that the horizontally and vertically polarized photons act as the pump and probe field for the molecules, respectively. The probe field transmitted through the sample is frequency-dispersed, and the change in the transmitted photon number is registered as a function of the frequency $\omega$, pump frequency $\omega_\p$, and external delay $\dt$, yielding the signal $S(\omega,\dt ;\omega_\p)$.

The Hamiltonian used to describe this pump--probe process is written as $\hat{H} = \hat{H}_{\rm mol} + \hat{H}_{\rm field} + \hat{H}_{\rm mol-field}$. The first term gives the Hamiltonian of the photoactive degrees of freedom (DOFs) in molecules. 
The second term describes the free electric field.
In this work, the electronic ground state $\lvert 0 \rangle$, single-excitation manifold $\{\lvert e_\alpha \rangle\}$, and double-excitation manifold $\{\lvert f_{\bar\gamma} \rangle\}$ are considered as photoactive DOFs. The overline of the subscripts indicates the state in the double-excitation manifold. The optical transitions are described by the dipole operator $\hat{\bmu} = \hat{\bmu}_+ + \hat{\bmu}_-$, where $\hat{\bmu}_- = \sum_\alpha \bmu_{\alpha 0} \lvert 0 \rangle \langle e_\alpha \rvert + \sum_{\alpha \bar\gamma} \bmu_{\bar\gamma \alpha} \lvert e_\alpha \rangle\langle f_{\bar\gamma} \rvert$ and $\hat{\bmu}_+ = [\hat{\bmu}_-]^\dagger$. The rotating-wave approximation enables the expression of the molecule-field interaction as $\hat{H}_{\rm mol-field}(t) = -\hat{\bmu}_- \cdot \hat{\bE}^+(t) -\hat{\bmu}_+ \cdot \hat{\bE}^-(t)$. The signal is expressed as \cite{Dorfman:2016da} 
\begin{align}
	S(\omega, \dt ;\omega_\p)
	=
	\Im 
	\int^\infty_{-\infty}dt\,
	e^{i\omega t} 
	\tr
	[ \hat{\bE}_\V^-(\omega) \cdot \hat{\bmu}_- \hat{\rho}(t) ],
	\label{eq:signal}
\end{align}
where $\hat\rho(-\infty) = \lvert 0 \rangle\langle 0 \rvert \otimes \lvert \psi_\twin \rangle\langle \psi_\twin \rvert$ and $\hat{\bE}_\lambda^-(\omega)=\int_0^\infty dt\, \hat{\bE}_\lambda^-(t) e^{-i\omega t}$.
We expand the density operator $\hat\rho(t)$ with respect to $\hat{H}_{\rm mol-field}$ to the third order, resulting in the sum of eight contributions classified as stimulated emission (SE), ground-state bleaching (GSB), excited-state absorption (ESA), and double quantum coherence (DQC). 
The DQC signal decays rapidly in comparison with the others when $\dt$ is sufficiently long compared to the timescale of environmental reorganization (see Section~S1 of Supplementary Material for details); hence, the DQC is disregarded in this work. Each contribution is expressed as follows:
\begin{align}
	S_{x,y}&(\omega,\dt ;\omega_\p)
\notag \\
	&=
	\sum_{\lambda_n = \H, \V }
	\Im \int^\infty_{-\infty} dt\, e^{i\omega t}
	\iiint^\infty_0 d^3 s\,
	R_{x,y}^{\V\lambda_3\lambda_2\lambda_1}
	(s_3,s_2,s_1)
\notag \\
    &\quad \times
	C_{x,y}^{\V\lambda_3\lambda_2\lambda_1}
	(\omega,t;s_3,s_2,s_1),
	\label{eq:signal-contribution}
\end{align}
where $x$ indicates rephasing (r) or non-rephasing (nr), and $y$ indicates the GSB, SE, or ESA.
Here, $R_{x,y}^{\lambda_4\lambda_3\lambda_2\lambda_1}(s_3,s_2,s_1)$ and $C_{x,y}^{\lambda_4\lambda_3\lambda_2\lambda_1}(\omega,t;s_3,s_2,s_1)$ are the third-order response functions of the molecules and four-body correlation functions of the field operators, respectively.

For demonstration purposes, we focused on rephasing the SE contribution. Details of the GSB and ESA are provided in Section~S2 of Supplementary Material. The rephasing SE contribution is given by
$
	R_{\rephasing , \SE} 	
	^{\lambda_4\lambda_3\lambda_2\lambda_1}
	(s_3,s_2,s_1)
	=
	(i/\hbar)^3
	\sum_{\alpha\beta\gamma\delta}
	\langle 
		\mu_{\delta 0}^{\lambda_4} 
		\mu_{\gamma 0}^{\lambda_3}
		\mu_{\beta  0}^{\lambda_2}
		\mu_{\alpha 0}^{\lambda_1} 
	\rangle 
	G_{\gamma 0}(s_3)
	G_{\gamma\delta\gets\alpha\beta}(s_2)
	G_{0\beta}(s_1)
$.
Here, we have defined 
	$\langle 
		\mu_{\epsilon\zeta }^{\lambda_4} 
		\mu_{\gamma  \delta}^{\lambda_3} 
		\mu_{\beta   0     }^{\lambda_2} 
		\mu_{\alpha  0     }^{\lambda_1} 
	\rangle 
	= 
	\langle 
		(\bmu_{\epsilon\zeta }\cdot\be_{\lambda_4})
		(\bmu_{\gamma  \delta}\cdot\be_{\lambda_3}) 
		(\bmu_{\beta   0     }\cdot\be_{\lambda_2})
		(\bmu_{\alpha  0     }\cdot\be_{\lambda_1})  
	\rangle_{\rm ori}$, 
where the brackets $\langle\dots\rangle_{\rm ori}$ represent the average over molecular orientations \cite{SchlauCohen:2011cw,Schlawin:2022po}. 
The matrix element of the time-evolution operator $G_{\gamma\delta\gets\alpha\beta}(t)$ is defined by $\rho_{\gamma\delta}(t) = \sum_{\alpha\beta} G_{\gamma\delta\gets\alpha\beta}(t) \rho_{\alpha\beta}(0)$, and $G_{\alpha\beta}(t)$ is the abbreviation of $ G_{\alpha\beta\gets\alpha\beta}(t)$.
By substituting the response function and field correlation function \cite{Ishizaki:2020jl} into Eq.~\eqref{eq:signal-contribution}, we obtain the following SE signal:
\begin{align}
	S_{\rephasing, \SE}(\omega,\dt ;\omega_\p)
	&=
	-
	\eta
	\Phi(\omega-\bar\omega_\V)
	\Re
	\sum_{\alpha\beta\gamma\delta}
	\sum_{\lambda_n=\H,\V}
	\langle 
		\mu_{\delta 0}^{\lambda_4} 
		\mu_{\gamma 0}^{\lambda_3}
		\mu_{\beta  0}^{\lambda_2}
		\mu_{\alpha 0}^{\lambda_1} 
	\rangle 
\notag \\
    &\quad \times
	G_{\gamma 0}[\omega]
	F_{\gamma\delta\gets\alpha\beta}
	^{{\lambda_4\lambda_3\lambda_2\lambda_1}}
	(\omega,\dt;0)
	G_{0\beta}[\omega_\p - \omega]
\notag \\
    &\quad
	+
	S^{\mathrm{(c)}}_{\rephasing,\SE}(\omega),
	\label{eq:rephasing-SE}
\end{align}
where $\eta = \zeta^2 \calE(\bar\omega_\H)^2 \calE(\bar\omega_\V)^2/\hbar^3$.
The second term in Eq.~\eqref{eq:rephasing-SE} originates from a field commutator. This term does not depend on $\dt$. Therefore, the contribution to the signal can be ignored by considering the difference spectrum:
\begin{align}
	\Delta S(\omega,\dt ;\omega_\p) = S(\omega,\dt ;\omega_\p) - S(\omega,\dt=0 ;\omega_\p).
	\label{eq:difference-spectrum}
\end{align}
The Fourier--Laplace transform of $G_{\alpha\beta}(t)$ is written as $G_{\alpha\beta}[\omega]$, and $F_{\gamma\delta\gets\alpha\beta}^{{\lambda_4\lambda_3\lambda_2\lambda_1}}(\omega,\dt;s_1)$ is defined as 
\begin{multline}
	F_{\gamma\delta\gets\alpha\beta}
	^{{\lambda_4\lambda_3\lambda_2\lambda_1}}
	(\omega,\dt;s_1)
	=
	\int^\infty_0 ds_2 \,G_{\gamma\delta\gets\alpha\beta}(s_2)
	e^{-i(\omega-\bar\omega_\V) \dt}
\\
	\times
	[
		D_1(s_2+s_1-\dt) 
		e^{i(\omega-\bar\omega_\V)(s_2+s_1)}
		\delta_{\lambda_1\H}
		\delta_{\lambda_2\H}
		\delta_{\lambda_3\V}
		\delta_{\lambda_4\V}
\\
		+
		D_1(s_2+s_1+\dt) 
		e^{i(\omega-\bar\omega_\H)(s_2+s_1)}
		\delta_{\lambda_1\H}
		\delta_{\lambda_2\V}
		\delta_{\lambda_3\H}
		\delta_{\lambda_4\V}
	],
	\label{eq:F-func}
\end{multline}
where $D_n(t) = (2\pi)^{-1}\int^\infty_{-\infty}d\omega\,e^{-i\omega t}[\Phi(\omega)]^n$.
As discussed in Ref.~\citenum{Ishizaki:2020jl}, Eq.~\eqref{eq:F-func} for $\dt > T_\e/2$ can be simplified as 
\begin{align}
	F_{\gamma\delta\gets\alpha\beta}	
	^{{\lambda_4\lambda_3\lambda_2\lambda_1}}(\omega,\dt;0) 
	\propto 
	G_{\gamma\delta\gets\alpha\beta}(\dt) 
	\delta_{\lambda_1\H} \delta_{\lambda_2\H} 	
	\delta_{\lambda_3\V} \delta_{\lambda_4\V},
	\label{eq:F-func-limit}
\end{align} 
which is independent of $\omega$. 
To obtain the information contents of the signal, we assume that the time evolution in the $t_1$ and $t_3$ periods is described as $G_{\alpha\beta}(t)=e^{-(i\omega_{\alpha\beta}+\epsilon_+)t}$, thereby leading to the expression of the rephasing SE signal, $G_{\gamma 0}[\omega]
F_{\gamma\delta\gets\alpha\beta}^{{\lambda_4\lambda_3\lambda_2\lambda_1}}(\omega,\dt;0)
G_{0\beta}[\omega_\p - \omega]
\propto 
\delta(\omega -\omega_{\gamma 0})
\delta(\omega_\p - \omega - \omega_{\beta 0})
G_{\gamma\delta\gets\alpha\beta}(\dt)$. It can be understood that the non-classical correlation between the entangled photon pair generated via the PDC pumped with a monochromatic laser restricts the possible optical transitions, ($0 \to e_\beta$, $0 \to e_\gamma$), for a given pump frequency. Therefore, Eqs.~\eqref{eq:rephasing-SE} and \eqref{eq:F-func-limit} indicate that the state-to-state dynamics in the molecules are temporally resolved by sweeping the external delay $\dt$ in the time region longer than half of the entanglement time, $T_\e/2$.
It should be mentioned that phenomena similar to the specific selective excitation described above have been discussed in Ref.~\citenum{Schlawin:2013dq} in the context of manipulation of two-excitation distributions by the non-classical photon correlation.

Notably, in the limit $T_\e\to 0$, the third-order signal in Eq.~\eqref{eq:signal} corresponds to the spectral information along the anti-diagonal line of the absorptive 2D spectrum obtained using the photon-echo technique in the impulsive limit:
\begin{align}
	S(\omega,\dt;\omega_\p)
	=
	-\mathcal{S}_{\rm 2D}(\omega, \dt, \omega_\p -\omega),
	\label{eq:correspondence}
\end{align} 
except for the $\dt$-independent term \cite{Ishizaki:2020jl}, as shown in Fig.~\ref{fig:2} (the explicit expression of the 2D photon-echo spectrum, $\mathcal{S}_{\rm 2D}(\omega_3, \dt, \omega_1)$, is given in Appendix~\ref{app:secB}).
It is noted that the sign of the quantum spectrum, $S(\omega,\dt;\omega_\p)$, is the opposite of the sign of the classical 2D photon-echo spectrum, $\mathcal{S}_{\rm 2D}(\omega, \dt, \omega_\p -\omega)$, as shown in Eq.~\eqref{eq:correspondence}.
In the following numerical results, the spectrum $S(\omega,\dt;\omega_\p)$ is plotted multiplied by a minus sign for clarity.
Equation~\eqref{eq:correspondence} also indicates that the pump-probe signal shows no collective two-particle contributions (see Section~S3 of Supplementary Material for details). This result is consistent with the arguments in Refs.~\citenum{Muthukrishnan:2004in,Richter:2011co}.

Furthermore, the phase matching function $\Phi(\omega-\bar\omega_\V) = \sinc[(\omega-\bar\omega_\V)T_\e/2]$ in Eq.~\eqref{eq:rephasing-SE} can selectively enhance a specific spectral region of the signal by varying the center frequency $\bar\omega_\V$ of the generated beam. The width at half maximum of $\Phi(\omega-\bar\omega_\V)$ is approximately given by $\lvert \omega - \bar\omega_\V \rvert \simeq 4/T_\e$. 
Interestingly, this corresponds to a \textit{sinc filter} in signal processing \cite{Owen:2007pr}.
Therefore, the entanglement time $T_\e$ plays a dual role of the knob for controlling the accessible time region of the dynamics in molecules, $\dt > T_\e /2$, and the degree of spectral selectivity, $\lvert \omega - \bar\omega_\V \rvert \simeq 4/T_\e$.
It is noted that similar spectral filtering can be realized with classical light \cite{Oshea:2001hi}, and has been utilized for selective excitation in multidimensional spectra \cite{Tollerud:2014is}.

\section{Numerical results}

In the following, we discuss roles of the entanglement time on the temporal resolution and spectral selectivity through numerical investigations of the signals in Eqs.~\eqref{eq:rephasing-SE}--\eqref{eq:F-func} and Eqs.~(S7)--(S11) of the Fenna-Matthews-Olson (FMO) pigment-protein complex in the photosynthetic green sulfur bacterium \textit{Chlorobium tepidum} \cite{Li:1997cr,CamaraArtigas:2003ch,Tronrud:2009ch} (Fig.~\ref{fig:1}C).
Due to its relatively small size, it has been widely studied experimentally and theoretically as a prototypical system for discussing photosynthetic energy transfer using nonlinear optical spectroscopy \cite{Freiberg:1997ex,Brixner:2005wu,Engel:2007hb,Fujihashi:2015kz}.
Our model includes seven single-excitation states $\{e_1, \cdots, e_7 \}$ and 21 double-excitation states $\{ f_{\bar{1}}, \cdots, f_{\bar{21}} \}$ (for details on the model, see Appendix~\ref{app:secA}).
The matrix elements $G_{\alpha\beta}[\omega]$ and $G_{\gamma\delta\gets\alpha\beta}(t)$ in Eq.~\eqref{eq:rephasing-SE} and Eqs.~(S7)--(S11) are calculated using the cumulant expansion for the fluctuations in the electronic energies and the modified Redfield theory \cite{Zhang:1998eo} (see Appendix~\ref{app:secA}).

%%%%%%%%%%
\subsection{Limit of short entanglement time}

We investigate the correspondence between the classical 2D Fourier-transformed photon-echo signal and the transmission signal with entangled photons.
Figure~\ref{fig:3}A presents the difference spectra, $\Delta S(\omega,\dt;\omega_\p)$, with quantum-entangled photon pairs in the limit of $T_\e\to 0$.
The waiting times are $\dt = 0.5\,{\rm ps}$ and $2\,{\rm ps}$.
The temperature was set as $T=77\,{\rm K}$.
For comparison, we depict the 2D photon-echo spectra, $\mathcal{S}_{\rm 2D}(\omega_3, \dt, \omega_1)$, generated by four laser pulses in the impulsive limit in Fig.~\ref{fig:3}B.
For the calculations, we chose the HHVV sequence for the polarizations of the four laser pulses so that the polarization sequence was the same as that of the entangled photon pair in the limit of short entanglement time.
Figure~\ref{fig:3}B shows six separate peaks at positions marked by black squares.
The diagonal peaks centered in the vicinity of $(\omega_1,\omega_3 )=(\epsilon_m,\epsilon_m)$ are labeled as DP$m$, whereas the cross-peaks located around $(\omega_1,\omega_3 )=(\epsilon_m,\epsilon_n)$ are labeled as CP$mn$.
As can be seen in Eq.~\eqref{eq:correspondence} and Fig.~\ref{fig:2}, in the difference spectra, peaks corresponding to DP$m$ and CP$mn$ appear near $(\omega_\p,\omega )=(2\epsilon_m,\epsilon_m)$ and $(\omega_\p,\omega )=(\epsilon_m + \epsilon_n,\epsilon_n)$, respectively.
Thus, each of the six peaks at the positions indicated by the black square in Fig.~\ref{fig:3}A shows the spectral information of the peak at the same label position in Fig.~\ref{fig:3}B.
It is noted that from the definition of the difference spectrum in Eq.~\eqref{eq:difference-spectrum} the decay of the SE signal at finite delay times $\dt$ appears as a negative signal, as presented by DP2 and DP5 in Fig.~\ref{fig:3}A.

While the cross-peaks at $\dt=0$ indicate coupled excited states, the appearance of the cross-peaks with increasing waiting time, $\dt$, indicates a relaxation process from a higher exciton state to a lower exciton state.
As time progressed, the appearance of CP51 can be observed in Fig.~\ref{fig:3}.
This behavior reflects the $e_5 \to e_1$ relaxation process, as presented in Supporting information, Fig.~S2.
Similarly, the increase in the peak amplitude of CP21 during $\dt$ was attributed to the $e_2 \to e_1$ relaxation process.
However, the Liouville pathways involving $e_3$, $e_4$, $e_6$, and $e_7$ states have much smaller amplitudes than CP21 and CP51.
Hence, it is difficult to extract information on the energy transfer processes involving these excitation states owing to spectral congestion.

%%%%%%%%%%%
\begin{figure}
    \centering
    \includegraphics{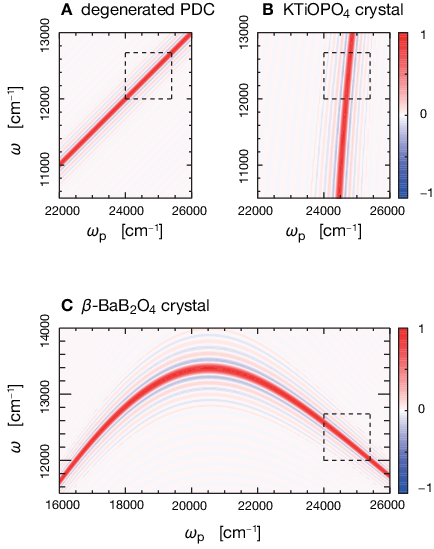}
    \caption{
	Two-dimensional plots of the phase-matching function $\sinc[\Delta k(\omega_\p -\omega,\omega)L/2]$ as a function of $\omega$ and $\omega_\p$ for cases of (A) degenerated PDC for the case of $T_\e=500\,{\rm fs}$, (B) periodically poled $\mathrm{KTiOPO_4}$ (PPKTP) crystal at $T_{\rm KTP}=323\,{\rm K}$ and $\Lambda=9.75\,{\rm \mu m}$ for the case of $T_\e = 10\,{\rm fs}$ (corresponding to $L = 0.15\,{\rm mm}$), and (C) $\beta$-$\mathrm{BaB_2O_4}$ (BBO) crystal at $\theta=41.7^\circ$ for the case of $T_\e = 500\,{\rm fs}$ (corresponding to $L = 2.8\,{\rm mm}$).
	In panel~(A), the phase-matching function was computed using Eq.~\eqref{eq: phase-matching-func} under the degeneracy condition ($\bar\omega_\H = \bar\omega_\V = \omega_\p /2$).
	The phase-matching function of panels~(B) and (C) were calculated with the Sellmeier equations in Refs.~\citenum{Fradkin:1999tu,Konig:2004ex} and Ref.~\citenum{Kato:1986se}, respectively.
	It is noted that the refractive index given by the Sellemeier equation for the BBO crystal is assumed to be independent of the crystal's temperature, in contrast to the case of the PPKTP crystal (see Section~S7 of Supplementary Material).
	The entanglement times in the BBO crystal (the PPKTP crystal) were evaluated by calculating the central frequencies and group velocities of the generated twin photons when pumped at $24000\,{\rm cm}^{-1}$ ($24700\,{\rm cm}^{-1}$).
	In each panel, the dashed-line box indicates the spectral range of the FMO complex ($24000\,{\rm cm}^{-1} < \omega_\p < 25400\,{\rm cm}^{-1}$, $12000\,{\rm cm}^{-1} < \omega < 12700\,{\rm cm}^{-1}$).
 }
    \label{fig:4}
\end{figure}
%%%%%%%%%%%%

%%%%%%%%%%%
\begin{figure}
    \centering
    \includegraphics{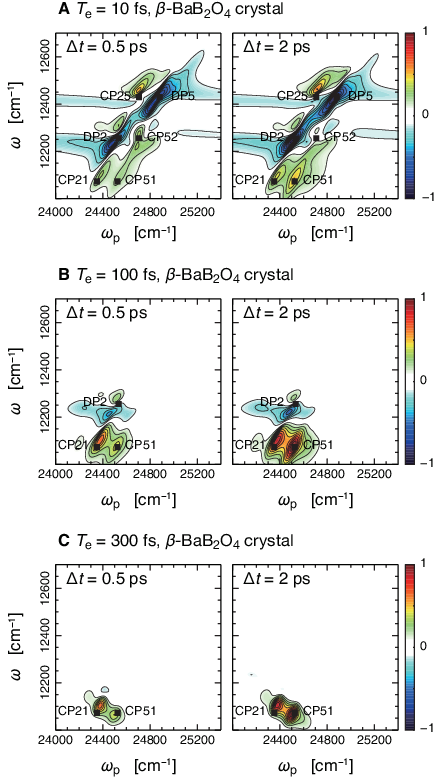}
    \caption{
	Difference Spectra, $\Delta S(\omega,\dt;\omega_\p)$, as a function of $\omega$ and $\omega_\p$ with entangled photon pairs generated via the BBO crystal for (A) $T_\e = 10\,{\rm fs}$ (i.e., $L = 0.056\,{\rm mm}$), (B) $T_\e = 100\,{\rm fs}$ ($L = 0.56\,{\rm mm}$), and (C) $T_\e = 300\,{\rm fs}$ ($L = 1.68\,{\rm mm}$).
	The entanglement time was evaluated by calculating the central frequencies and group velocities of twin photons when pumped at $24000\,{\rm cm}^{-1}$.
	The propagation angle of the beam with respect to the optic axis is set to $\theta=41.1^\circ$.
	The scattering angles are set to $\phi_1=\phi_2=0$.
	The other parameters are the same as that in Fig.~\ref{fig:3}.
	The normalization of contour plots~(A)--(C) is such that the maximum value of each spectrum at $\dt = 2\,{\rm ps}$ is unity, and equally spaced contour levels ($\pm 0.1$, $\pm 0.2$, \dots) are drawn.
 }
    \label{fig:5}
\end{figure}
%%%%%%%%%%%%

%%%%%%%%%%%
\begin{figure}
    \centering
    \includegraphics{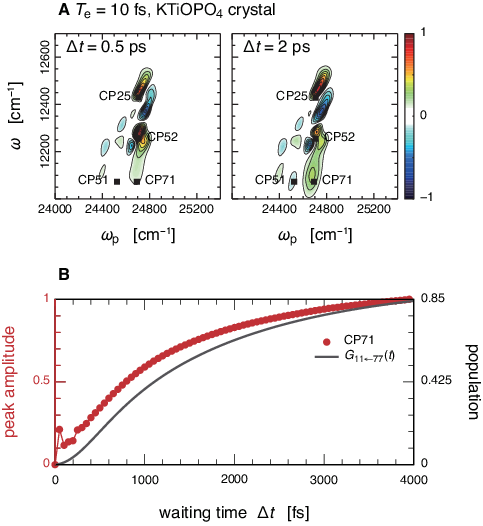}
    \caption{
	(A) Difference Spectra, $\Delta S(\omega,\dt;\omega_\p)$, as a function of $\omega$ and $\omega_\p$ with entangled photon pairs generated via the PPKTP crystal for $T_\e = 10\,{\rm fs}$ (i.e., $L = 0.15\,{\rm mm}$).
	The entanglement time was evaluated by calculating the central frequencies and group velocities of the generated twin photons when pumped at $24700\,{\rm cm}^{-1}$.
	The poling period and the crystal's temperature are set to $\Lambda=2.47\,{\rm \mu m}$ and $T_{\rm KTP}=323\,{\rm K}$.
	The normalization of contour plot~(A) is such that the maximum value of each spectrum at $\dt = 2\,{\rm ps}$ is unity, and equally spaced contour levels ($\pm 0.1$, $\pm 0.2$, \dots) are drawn.
	(B) Time evolution of the amplitudes of CP71 ($\omega_\p = 24692.5\,{\rm cm}^{-1}$, $\omega = 12072.5\,{\rm cm}^{-1}$).
    For comparison, the black line represents the matrix element of the time-evolution operator, $G_{11\gets 77}(t)$, calculated directly in the modified Redfield theory, which corresponds to $e_7 \to e_1$ transport.
    The normalization of the amplitude of CP71 is such that the maximum value of the peak amplitude is unity.
    }
    \label{fig:6}
\end{figure}
%%%%%%%%%%%%

%%%%%%%%%%
\subsection{Cases of finite entanglement times}

To discuss the roles of the entanglement time in the temporal resolution and spectral selectivity, we investigated cases of nondegenerate down-conversion, $\bar\omega_\H \ne \bar\omega_\V$. 
A nondegenerate type-II PDC experiment in the visible frequency region is possible, for example, using a periodically poled $\mathrm{KTiOPO_4}$ (PPKTP) crystal \cite{Kim:2006ph,Fedrizzi:2007wa} and $\beta$-$\mathrm{BaB_2O_4}$ (BBO) crystal \cite{Yabushita:2004hy}, as shown in Fig.~\ref{fig:4}.

We first considered the nondegenerate PDC through the BBO crystal in the collinear configuration, where the scattering angles shown in Fig.~\ref{fig:1}A are set to $\phi_1=\phi_2=0$.
The phase-matching condition of a uniaxial birefringent crystal, such as a BBO crystal, depends on the propagation angle $\theta$ of the pump beam with respect to the optical axis (see Section~S4 of Supplementary Material).
Thus, the central frequencies of the generated twin beams can be tuned by changing the value of $\theta$.
Figure~\ref{fig:5} presents the difference spectra, $\Delta S(\omega,\dt;\omega_\p)$, as a function of $\omega$ and $\omega_\p$ with entangled photon pairs generated via the BBO crystal for (A) $T_\e = 10\,{\rm fs}$, (B) $T_\e = 100\,{\rm fs}$, and (C) $T_\e = 300\,{\rm fs}$.
The angle of the pump beam with respect to the optical axis is set to $\theta=41.1^\circ$, where the central frequencies of the twin photons nearly resonate with the pair of optical transitions ($0 \to e_5$ and $e_1 \to 0$). 
The other parameters were the same as those shown in Fig.~\ref{fig:3}A.
The value of the entanglement time in Fig.~\ref{fig:5} was evaluated by computing the group velocities of twin photons generated when pumped at $24000\,{\rm cm}^{-1}$.
Strictly speaking, the entanglement time depends on the value of the pump frequency $\omega_\p$ because the group velocities $v_\H$ and $v_\V$ depend on the value of $\omega_\p$.
However, the influence of this correction is small because the group velocities vary rather slowly far from the absorption resonances of the nonlinear crystal (see Supplementary Material, Fig.~S1).
The signal in Fig.~\ref{fig:5}A appears to be similar to that in Fig.~\ref{fig:3}A, which can be regarded as the limit for the short entanglement time.
As pointed out above, the spectral distribution of the phase-matching function can selectively enhance a specific region of the spectra allowed by the bandwidth $\lvert \omega - \bar\omega_\V \rvert \simeq 4/T_\e$.
This can be observed in Figs.~\ref{fig:5}B and C, where the intensities of the peaks except for CP21 and CP51 are suppressed with increasing entanglement time.
Simultaneously, the overall behavior of the signal at CP51 in Fig.~\ref{fig:5} is similar to the dynamics of the $e_5 \to e_1$transport, as demonstrated in Supplementary Material, Fig.~S3. It is because it is possible to extract relevant information on the excited-state dynamics from the signal in the time region longer than half of the entanglement time.

The results in Fig.~\ref{fig:5} suggest that the spectral filter of the phase-matching function can be used to extract information on weak signal contributions masked by strong peaks.
To demonstrate this ability, we investigated the case of a nondegenerate PDC using the PPKTP crystal.
In the PPKTP crystal, it is possible to tune the central frequencies of the generated twin beams by changing the values of the poling period and crystal temperature (see Section~S7 of Supplementary Material).
Figure~\ref{fig:6}A shows the difference spectra, $\Delta S(\omega,\dt;\omega_\p)$, as a function of $\omega$ and $\omega_\p$ with the entangled photon pairs generated via the PPKTP crystal for $T_\e = 10\,{\rm fs}$.
The poling period and crystal temperature were set to $\Lambda=2.47\,{\rm \mu m}$ and $T_{\rm KTP}=323\,{\rm K}$, respectively, where the central frequencies of the twin photons nearly resonate with the pair of optical transitions ($0 \to e_7$, $e_1 \to 0$).
The entanglement time was evaluated by calculating the central frequencies and group velocities of the generated twin photons when pumped at $24700\,{\rm cm}^{-1}$.
The other parameters were the same as those shown in Fig.~\ref{fig:3}A.
Figure~\ref{fig:6}B also shows the time evolution of the CP71 amplitude ($\omega_\p = 24692.5\,{\rm cm}^{-1}$ and $\omega = 12072.5\,{\rm cm}^{-1}$), as shown in Fig.~\ref{fig:6}A.
For comparison, the black line shown in Fig.~\ref{fig:6}B represents the matrix element of the time-evolution operator, $G_{11\gets 77}(t)$, which is calculated directly in the modified Redfield theory and corresponds to $e_7 \to e_1$ transport.
The intensities of CP21 and CP51 shown in Fig.~\ref{fig:6}A are suppressed because of the narrow spectral filter of the phase-matching function.
Consequently, the signal at CP71 in Fig.~\ref{fig:6}A is not distorted by interference with nearby cross-peak components, such as CP21 and CP51, and is better resolved compared with that in Fig.~\ref{fig:3}A.
Simultaneously, the overall behavior of the signal at CP71 in Fig.~\ref{fig:6}B is similar to the dynamics of $e_7 \to e_1$ transport.
Therefore, Fig.~\ref{fig:6} demonstrates that the spectral filter of the phase-matching function can be used to selectively enhance specific peaks within the congested 2D spectra of the FMO complex by controlling the entanglement time and central frequencies of the entangled photons while maintaining the ultrafast temporal resolution.

As indicated in Figs.~\ref{fig:5} and \ref{fig:6}, in the case of finite entanglement time, the role of the entanglement time is substantially equivalent to the temporal width of the classical laser pulse. In this sense, entangled photon pairs do not provide simultaneous improvement in temporal and frequency resolution over spectroscopy using classical laser pulses. However, it is interesting that the non-classical correlation between the twin photons enables the selective excitations of specific single-excitation states although a simple optical system and monochromatic laser are employed. This is one aspect of the usefulness of non-classical photon correlation for spectroscopic measurements. This insight encourages us to envision that employing more intricately engineered quantum states of light could expand the applicability of quantum light spectroscopy and molecular quantum metrology.

%Discussions%%%%%%%%%%%%%%%%%%%%%%%%%%%%
\section{Discussion}

In this work, we explored the roles of entanglement time for temporal resolution and spectral selectivity through numerical investigations of the entangled photon spectroscopy of FMO complexes.
The frequency-dispersed transmission measurement with entangled photons considered in our study exhibits three interesting features. First, nonclassical photon correlation enables time-resolved spectroscopy with monochromatic pumping \cite{Ishizaki:2020jl}.
The temporal width of the entangled photon pair is determined by the entanglement time, and hence can be tuned by the crystal length.
For example, as illustrated in Fig.~\ref{fig:5}A, the temporal width of the entangled photon pairs generated with the BBO crystal with a thickness of $0.056\,{\rm mm}$ was a few femtoseconds.
Therefore, transmission measurement can temporally resolve the dynamic processes of molecular systems occurring at femtosecond timescales without requiring sophisticated control of the temporal delay between femtosecond laser pulses.
Second, the spectral distribution of the phase-matching function can function as a frequency filter, which removes all optical transitions that fall outside the spectral width. 
The spectral distribution of the phase-matching function can be manipulated by changing the crystal length and phase-matching angle. 
Thus, specific peaks in crowded 2D spectra can be selectively enhanced or suppressed by controlling the phase-matching function, as in laser spectroscopic experiments using narrow-band pulses.
As demonstrated in Fig.~\ref{fig:5}, the selective enhancement of specific peaks in the congested 2D spectrum of the FMO complex can be achieved using the phase-matching function of the PDC process in BBO crystals with crystal lengths ranging from $0.056\,{\rm mm}$ to $1.68\,{\rm mm}$, which has been used experimentally \cite{Branning:2000si,Dayan:2004kg,Yabushita:2004hy,Lee:2006id,Eshun:2021in}.
Therefore, spectral filtering is feasible using current quantum optical techniques for generating entangled photons.
Third, the spectral distribution of the phase-matching function strongly depends on the properties of the nonlinear crystal. As shown in Figs.~\ref{fig:5} and \ref{fig:6}, the spectral filter effects can easily be adjusted by changing the nonlinear crystals and/or their properties.
Although we considered only BBO and PPKTP crystals in this study, there is a wide range of nonlinear crystals that have been used for PDC in the near-infrared and visible regions \cite{Dmitriev:2013ha}.
Therefore, we anticipate that transmission measurement can be applied not only to FMO complexes but also to other photosynthetic pigment-protein complexes such as the photosystem II reaction center \cite{Romero:2014jm,Fuller:2014iz,Fujihashi:2018in} by finding an appropriate nonlinear crystal corresponding to the spectral range of the molecular system of interest.

The feasibility of entangled two-photon spectroscopy was recently questioned by several research groups \cite{Landes:2021ex,Parzuchowski:2021se}.
The quantitative estimates using the upper bound on the isolated-entangled-pair cross section indicate that realistic sample concentrations event rates are orders of magnitude below the detection threshold of typical photon-counting systems \cite{Raymer:2021en}.
The same difficulties are expected to be faced in the case of the transmission measurement considered in our study.
One solution to overcome vanishingly weak nonlinear signals is to use high-gain squeezed vacuum states \cite{Cutipa:2022br}.
As demonstrated in Section~S8 of Supplementary Material, at least when the entanglement time is sufficiently short compared with characteristic timescales of the dynamics under investigation, the transmission measurement is capable of temporally resolving the excitation dynamics with high-gain squeezed vacuum states. Whether the measurement can be performed under other parameter conditions such as for long entanglement times is a subject for future research.

In the present work, we did not concentrate on the exploitation of the polarization control of entangled photon pairs. 
Polarization-controlled measurements have been considered a beneficial technique for separating crowded 2D spectra \cite{Hochstrasser:2001vp,Zanni:2001ks,Dreyer:2003th,SchlauCohen:2012dn,Westenhoff:2012fi}.
In this regard, exploiting the polarization of entangled photons has the potential to improve the resolution of quantum spectroscopy further. Moreover, hyperentanglement in frequency-time and polarization can be generated by the type-II PDC process \cite{Kwiat:1997hy}. Further research on the use of such more elaborately controlled quantum states of light for optical spectroscopy may lead to the development of novel time-resolved spectroscopic measurements with precision and resolution beyond the limits imposed by the laws of classical physics.

\begin{acknowledgments}
This study was supported by JSPS KAKENHI (Grant Numbers~JP17H02946 and JP21H01052) and MEXT KAKENHI (Grant Number~JP17H06437) in Innovative Areas ``Innovations for Light-Energy Conversion,'' MEXT Quantum Leap Flagship Program (Grant Numbers~JPMXS0118069242 and JPMXS0120330644).
Y.F. and M.H. are grateful for the financial support from MEXT KAKENHI (Grant Number~JP20H05839) in Transformative Research Areas (A), ``Dynamic Exciton: Emerging Science and Innovation (20A201),'' and JST PRESTO (Grant Numbers~JPMJPR19G8 and JPMJPR18GA).
K.M. acknowledges support from JSPS KAKENHI (Grant Number~JP21K14481).
\end{acknowledgments}

\appendix

\section{\label{app:secA}Molecular system}
The molecular Hamiltonian is given by $\hat{H}_{\rm mol} = \hat{H}_{\rm ex} + \hat{H}_{\rm ex-env} + \hat{H}_{\rm env}$ \cite{Fujihashi:2015kz}: The first term is the electronic excitation Hamiltonian, $\hat{H}_{\rm ex}=\sum_m \hbar \Omega_m \hat{B}^\dagger_m \hat{B}_m + \sum_{m \neq n} \hbar J_{mn} \hat{B}^\dagger_m \hat{B}_n$, where $\hbar\Omega_m$ is the Franck--Condon transition energy of the $m$th pigment, $\hbar J_{mn}$ is the electronic coupling between the pigments, and the excitation creation operator $\hat{B}^\dagger_m$ is introduced for the excitation vacuum $\vert 0 \rangle$ such that $\vert m \rangle = \hat{B}^\dagger_m \vert 0 \rangle$ and $\vert mn \rangle = \hat{B}^\dagger_m \hat{B}^\dagger_n \vert 0 \rangle$.
We assume that the environmental DOFs can be treated as an ensemble of harmonic oscillators with $\hat{H}_{\rm env}=\sum_\xi \hbar \omega_\xi (\hat{p}_\xi^2+\hat{q}_\xi^2)/2$, where $\{\hat{q}_\xi \}$ are the dimensionless normal-mode coordinates and $\{ \omega_\xi \}$ and $\{\hat{p}_\xi \}$ are the corresponding frequencies and momenta, respectively.
The last term is the electronic-environmental interaction and is expressed as $\hat{H}_{\rm ex-env}=\sum_m \hat{u}_m \hat{B}^\dagger_m \hat{B}_m $, where $\hat{u}_m =-\sum_\xi \hbar \omega_\xi d_{m \xi} \hat{q}_\xi$, and $d_{m\xi}$ denotes the coupling constant between the $m$th pigment and the $\xi$th normal mode.
In the eigenstate representation, the excitation Hamiltonian can be written as $\hat{H}_{\rm ex}=\epsilon_0  \lvert 0 \rangle\langle 0 \rvert + \sum_\alpha \epsilon_\alpha \lvert e_\alpha \rangle\langle e_\alpha \rvert +\sum_{\bar\gamma} \epsilon_{\bar\gamma} \lvert  f_{\bar\gamma} \rangle\langle f_{\bar\gamma} \rvert$, where $\lvert e_\alpha \rangle=\sum_m V_{m\alpha} \hat{B}^\dagger_m \vert 0 \rangle$ and $\vert f_{\bar\gamma} \rangle = \sum_{mn} W_{mn,{\bar\gamma}} \hat{B}^\dagger_m \hat{B}^\dagger_n \vert 0 \rangle$.

Because $\{\hat{q}_\xi \}$ are normal mode coordinates, the dynamics of $\hat{u}_m (t)=e^{i \hat{H}_{\rm env} t/\hbar} \hat{u}_m e^{-i \hat{H}_{\rm env} t/\hbar}$ can be described as a Gaussian process \cite{Kubo:1985bs}.
By applying the second-order cumulant expansion to the fluctuations in the electronic energies, the third-order response function is expressed in terms of the line-broadening function, $g_m (t)=\int_0^t ds_1 \int_0^{s_1} ds_2 C_m (s_2) /\hbar^2$, where $C_m (t)=\langle \hat{u}_m (t) \hat{u}_m (0) \rangle$ is expressed as $C_m (t)=(\hbar/\pi) \int_0^\infty d\omega J_m (\omega)[\coth (\hbar\omega/2 k_{\rm B} T)\cos\omega t-i\sin\omega t]$ in terms of the spectral density, $J_m (\omega)$.
The third-order response function of the molecules was computed using cumulant expansion for the fluctuations in the electronic energies \cite{Zhang:1998eo}. 
In this study, the spectral density is modeled as $J_m(\omega)=32 \lambda_{\rm env} \gamma_{\rm env}^3 /(\omega^2 + 4\gamma_{\rm env}^2)^2$, where $\lambda_{\rm env}$ and $\gamma_{\rm env}^{-1}$ represent the energy and timescale of environmental reorganization, respectively \cite{Ishizaki:2020jy}.
The time evolution of the electronic excitations during the waiting time, $\dt$, was computed in the modified Redfield theory \cite{Zhang:1998eo,Yang:2002ik}.

The FMO complex is a trimer made of identical subunits, each containing eight bacteriochlorophyll {\it a} (BChl{\it a}) molecules \cite{Tronrud:2009ch}.
Because the eighth BChl is only loosely bound, this pigment is usually lost in the majority of the FMO complexes during the isolation procedure \cite{Tronrud:2009ch,SchmidtamBusch:2011ba}.
Therefore, we did not consider the eighth BChl concentration in this study.
The parameters in the molecular Hamiltonian were obtained from Refs.~\citenum{Brixner:2005wu,Cho:2005bm}. The atomic coordinates of the FMO complex were based on the X-ray crystallographic structure (PDB code:1M50) \cite{CamaraArtigas:2003ch}. 
The $\rm Q_y$ electric transition dipoles were assumed to be placed along the ${\rm N_B}$--$\rm N_D$ axis, and the electric dipole strength of monomeric BChl{\it a} is $28.7\,{\rm D}^2$ \cite{Abramavicius:2008un,Voronine:2008ch}. We set the reorganization energy and relaxation time to $\lambda_{\rm env}=55\,{\rm cm}^{-1}$ and $\gamma_{\rm env}^{-1}=100\,{\rm fs}$, respectively. 
We modeled static disorder by adding a Gaussian disorder (the standard deviation is $20\,{\rm cm}^{-1}$) for each diagonal term in $\hat{H}_{\rm ex}$.
These values were used to fit the experimental absorption and circular dichroism spectra of the FMO complex at $77\,{\rm K}$ \cite{Abramavicius:2008un,Voronine:2008ch}.

\section{\label{app:secB}Classical light}
We considered the heterodyned 2D photon echo signal generated by three laser pulses in the impulsive limit.
The signal is Fourier-transformed with respect to the time delay between the first and second pulses, $t_1$, and the time delay between the third and local oscillator pulses, $t_3$.
The Fourier-transform frequency variables conjugate to $t_1$ and $t_3$ are denoted as $\omega_1$ and $\omega_3$, respectively.
The 2D photon-echo spectrum is expressed as 
\begin{align}
    \mathcal{S}_{\rm 2D}(\omega_3, \dt, \omega_1)
    =
    \mathcal{S}_{\rm r}(\omega_3, \dt, \omega_1)+\mathcal{S}_{\rm nr}(\omega_3, \dt, \omega_1)
\end{align}
in terms of the rephasing and non-rephasing contributions
\begin{align}
    \mathcal{S}_{\rephasing}(\omega_3, \dt, \omega_1)
    &=
    \Im 
    \sum_{y} 
    \int_0^\infty dt_3 e^{i\omega_3 t_3}
    \int_0^\infty dt_1 e^{-i\omega_1 t_1}
\notag \\
    &\quad\times
    R_{\rephasing, y}	
	^{\lambda_4\lambda_3\lambda_2\lambda_1}
	(t_3,\dt,t_1),
\end{align}
\begin{align}
    \mathcal{S}_{\nonrephasing}(\omega_3, \dt, \omega_1)
    &=
    \Im 
    \sum_{y}
    \int_0^\infty dt_3 e^{i\omega_3 t_3}
    \int_0^\infty dt_1 e^{i\omega_1 t_1}
\notag \\
    &\quad\times
    R_{\nonrephasing, y}	
	^{\lambda_4\lambda_3\lambda_2\lambda_1}
	(t_3,\dt,t_1),
\end{align}
where $y$ indicates GSB, SE, or ESA.

%

% Supplementary Information %%%%%%%%%%%%%%%%%%%

\clearpage
\widetext
\begin{center}
\textbf{\large Supplementary Information: Probing exciton dynamics with spectral selectivity through the use of quantum entangled photons}
\end{center}

\setcounter{equation}{0}
\setcounter{figure}{0}
\setcounter{table}{0}
\setcounter{page}{1}
\makeatletter
\renewcommand{\theequation}{S\arabic{equation}}
\renewcommand{\thefigure}{S\arabic{figure}}
\renewcommand{\bibnumfmt}[1]{[S#1]}
\renewcommand{\citenumfont}[1]{S#1}

\begin{center}
  Yuta Fujihashi,${}^{1,2}$ Kuniyuki Miwa,${}^{3,4}$ Masahiro Higashi,${}^{1,2}$ and Akihito Ishizaki${}^{3,4}$\\
  \vspace{0.2cm}{\small $^1${\em Department of Molecular Engineering, Kyoto University, Kyoto 615-8510, Japan}\\
    ${}^2${\em PRESTO, Japan Science and Technology Agency, Kawaguchi 332-0012, Japan}\\
    ${}^3${\em Institute for Molecular Science, National Institutes of Natural Sciences, Okazaki 444-8585, Japan}\\
    ${}^4${\em Graduate Institute for Advanced Studies, SOKENDAI, Okazaki 444-8585, Japan}}
\end{center}

%%%%%%%%%%%%%%%%%%%%
\section*{S1. Double quantum coherence signal}

\begin{figure}
\centering
    \includegraphics{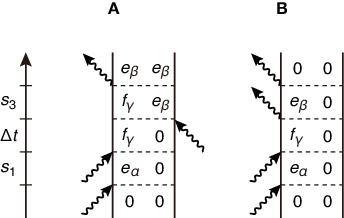}
    \caption{
    The two double-sided Feynman diagrams contributing to the DQC signal.}
    \label{fig:S4}
\end{figure}

In this section, we investigate the double quantum coherence (DQC) signal.
As displayed in Fig.~\ref{fig:S4}, there are two Liouville pathways contributing to the DQC signal.
In the following, we focus on the pathway in presented in Fig.~\ref{fig:S4}(A).
The DQC signal is expressed as follows:
\begin{align}
	S_{x,\DQC}(\omega,\dt ;\omega_\p)
	&=
	\sum_{\lambda_n = \H, \V }
	\Im \int^\infty_{-\infty} dt\, e^{i\omega t}
	\iiint^\infty_0 d^3 s\,
	R_{x,\DQC}^{\V\lambda_3\lambda_2\lambda_1}
	(s_3,s_2,s_1)
\notag \\
    &\quad \times
	C_{x,\DQC}^{\V\lambda_3\lambda_2\lambda_1}
	(\omega,t;s_3,s_2,s_1),
	\label{eq:DQC-contribution}
\end{align}
where
\begin{align}
	R_{x ,\DQC} 	
	^{\lambda_4\lambda_3\lambda_2\lambda_1}
	(s_3,s_2,s_1)
	=
	\left(\frac{i}{\hbar}\right)^3
	\sum_{\alpha\beta\gamma\delta}
	\langle 
		\mu_{\delta 0}^{\lambda_4} 
		\mu_{\gamma 0}^{\lambda_3}
		\mu_{\beta  0}^{\lambda_2}
		\mu_{\alpha 0}^{\lambda_1} 
	\rangle 
	G_{\gamma 0}(s_3)
	G_{\gamma\delta\gets\alpha\beta}(s_2)
	G_{0\beta}(s_1),
	\label{eq:response-DQC}
\end{align}
\begin{align}
	C_{x,\DQC}^{\lambda_4\lambda_3\lambda_2\lambda_1}
	(\omega,t;s_3,s_2,s_1)
	&=
	\langle
	\E^-_{\H}(t-s_3+\dt) \E^-_{\V}(\omega) \E^+_{\H}(t-s_3-s_2+\dt) \E^+_{\V}(t-s_3-s_2-s_1)
	\rangle
%	\,
\notag \\
	&\quad\times
	\delta_{\lambda_1\V}
	\delta_{\lambda_2\H}
	\delta_{\lambda_3\H}
	\delta_{\lambda_4\V}
\notag \\
	&\quad
	+
	\langle
	\E^-_{\H}(t-s_3+\dt) \E^-_{\V}(\omega) \E^+_{\V}(t-s_3-s_2) \E^+_{\H}(t-s_3-s_2-s_1+\dt)
	\rangle
%	\,
\notag \\
	&\quad\times
	\delta_{\lambda_1\H}
	\delta_{\lambda_2\V}
	\delta_{\lambda_3\H}
	\delta_{\lambda_4\V}.
	\label{eq:correlation-DQC}
\end{align}
In the limit of $T_\e \to 0$, Eq.~\eqref{eq:correlation-DQC} leads to
\begin{align}
	C_{x,\DQC}^{\lambda_4\lambda_3\lambda_2\lambda_1}
	(\omega,t;s_3,s_2,s_1)
	&=
	\int_0^\infty dT e^{-i\omega T}
	\delta(t-s_3+\dt-T)
	\delta(s_1-\dt)
	e^{i\bar\omega_\H(s_1+s_2)}
	e^{-i\bar\omega_\V(t-s_3-s_2-T)}
\notag \\
	&\quad\times
	\delta_{\lambda_1\H}
	\delta_{\lambda_2\V}
	\delta_{\lambda_3\H}
	\delta_{\lambda_4\V}.
	\label{eq:correlation-DQC-limit}
\end{align}
By substituting Eqs.~\eqref{eq:response-DQC} and \eqref{eq:correlation-DQC-limit} to Eq.~\eqref{eq:DQC-contribution}, we obtain
\begin{align}
	S_{x,\DQC}(\omega,\dt ;\omega_\p)
	&=
	-
	\frac{i}{\hbar^3}
	\sum_{\lambda_n = \H, \V }
	\Im \int^\infty_{-\infty} dt
	\int^\infty_0 d s_3
	e^{i\omega s_3}
	\int^\infty_0 d s_2
	e^{i\omega_\p s_2}
	e^{-i(\omega-\omega_\p)\dt}
	\sum_{\alpha\beta\gamma\delta}
	\langle 
		\mu_{\delta 0}^{\lambda_4} 
		\mu_{\gamma 0}^{\lambda_3}
		\mu_{\beta  0}^{\lambda_2}
		\mu_{\alpha 0}^{\lambda_1} 
	\rangle 
\notag \\
	&\quad\times
	G_{\gamma 0}(s_3)
	G_{\gamma\delta\gets\alpha\beta}(s_2)
	G_{0\beta}(\dt)
	\delta_{\lambda_1\V}
	\delta_{\lambda_2\H}
	\delta_{\lambda_3\H}
	\delta_{\lambda_4\V}.
\end{align}
When $\dt$ is sufficiently long compared to the timescale of environmental reorganization, $G_{0\beta}(\dt) \approx 0$.
Thus, the DQC contribution corresponding to the diagram in Fig.~\ref{fig:S4}(A) is negligibly small in comparison to the other Liouville pathways.
Similarly, the DQC signal contributions in Fig.~\ref{fig:S4}(B) are also understood.

%%%%%%%%%%%%%%%%%%%%
\section*{S2. Frequency-dispersed transmission signal with entangled photon pairs}

The contributions of the signals in Eq.~(5) of the main text is computed as follows:
\begin{align}
	S_{\rephasing, \SE}(\omega,\dt ;\omega_\p)
	&=
	-
	\eta
	\Phi(\omega-\bar\omega_\V)
	\Re
	\sum_{\alpha\beta\gamma\delta}
	\sum_{\lambda_n=\H,\V}
	\langle 
		\mu_{\delta 0}^{\lambda_4} 
		\mu_{\gamma 0}^{\lambda_3}
		\mu_{\beta  0}^{\lambda_2}
		\mu_{\alpha 0}^{\lambda_1} 
	\rangle 
\notag \\
    &\quad\times
	G_{\gamma 0}[\omega]
	F_{\gamma\delta\gets\alpha\beta}
	^{{\lambda_4\lambda_3\lambda_2\lambda_1}}
	(\omega,\dt;0)
	G_{0\beta}[\omega_\p - \omega]
	+
	S^{\mathrm{(c)}}_{\rephasing,\SE}(\omega),
	\label{eq:rephasing-SE2}
\end{align}
\begin{align}
    S_{\rephasing, \GSB}(\omega,\dt ;\omega_\p)
	&=
	-
	\eta
	\Phi(\omega-\bar\omega_\V)
	\Re
	\sum_{\alpha\beta}
	\sum_{\lambda_n=\H,\V}
	\langle 
		\mu_{\beta 0}^{\lambda_4} 
		\mu_{\beta 0}^{\lambda_3}
		\mu_{\alpha  0}^{\lambda_2}
		\mu_{\alpha 0}^{\lambda_1} 
	\rangle 
\notag \\
    &\quad\times
	G_{\beta 0}[\omega]
	F_{00\gets 00}
	^{{\lambda_4\lambda_3\lambda_2\lambda_1}}
	(\omega,\dt;0)
	G_{0\alpha}[\omega_\p - \omega]
	+
	S^{\mathrm{(c)}}_{\rephasing,\GSB}(\omega),
	\label{eq:rephasing-GSB}
\end{align}
\begin{align}
    S_{\rephasing, \ESA}(\omega,\dt ;\omega_\p)
    =
	\eta
	\Phi(\omega-\bar\omega_\V)
	\Re
	\sum_{\alpha\beta\gamma\delta \bar\epsilon}
	\sum_{\lambda_n=\H,\V}
	\langle 
		\mu_{\delta \bar\epsilon}^{\lambda_4} 
		\mu_{\gamma \bar\epsilon}^{\lambda_3}
		\mu_{\beta  0}^{\lambda_2}
		\mu_{\alpha 0}^{\lambda_1} 
	\rangle 
	G_{\bar\epsilon \delta}[\omega]
	F_{\gamma\delta\gets\alpha\beta}
	^{{\lambda_4\lambda_3\lambda_2\lambda_1}}
	(\omega,\dt;0)
	G_{0 \beta}[\omega_\p - \omega],
	\label{eq:rephasing-ESA}
\end{align}
\begin{align}
    S_{\nonrephasing, \SE}(\omega,\dt ;\omega_\p)
    &=
    -
	\eta
	\Phi(\omega-\bar\omega_\V)
	\Re
	\sum_{\alpha\beta\gamma\delta}
	\sum_{\lambda_n=\H,\V}
	\langle 
		\mu_{\delta 0}^{\lambda_4} 
		\mu_{\gamma 0}^{\lambda_3}
		\mu_{\beta  0}^{\lambda_2}
		\mu_{\alpha 0}^{\lambda_1} 
	\rangle 
\notag \\
    &\quad\times
	G_{\gamma 0}[\omega]
	\int^\infty_0 ds_1 e^{i(\omega_\p-\omega)s_1}
	F_{\gamma\delta\gets\alpha\beta}
	^{{\lambda_4\lambda_3\lambda_2\lambda_1}}
	(\omega,\dt;s_1)
	G_{\alpha 0}(s_1)
	+
	S^{\mathrm{(c)}}_{\nonrephasing,\SE}(\omega),
	\label{eq:nonrephasing-SE}
\end{align}
\begin{align}
    S_{\nonrephasing, \GSB}(\omega,\dt ;\omega_\p)
    &=
    -
	\eta
	\Phi(\omega-\bar\omega_\V)
	\Re
	\sum_{\alpha\beta}
	\sum_{\lambda_n=\H,\V}
	\langle 
		\mu_{\beta 0}^{\lambda_4} 
		\mu_{\beta 0}^{\lambda_3}
		\mu_{\alpha  0}^{\lambda_2}
		\mu_{\alpha 0}^{\lambda_1} 
	\rangle 
\notag \\
    &\quad\times
	G_{\beta 0}[\omega]
	\int^\infty_0 ds_1 e^{i(\omega_\p-\omega)s_1}
	F_{00\gets 00}
	^{{\lambda_4\lambda_3\lambda_2\lambda_1}}
	(\omega,\dt;s_1)
	G_{\alpha 0}[\omega_\p - \omega]
	+
	S^{\mathrm{(c)}}_{\nonrephasing,\GSB}(\omega),
	\label{eq:nonrephasing-GSB}
\end{align}
\begin{align}
    S_{\nonrephasing, \ESA}(\omega,\dt ;\omega_\p)
    &=
	\eta
	\Phi(\omega-\bar\omega_\V)
	\Re
	\sum_{\alpha\beta\gamma\delta \bar\epsilon}
	\sum_{\lambda_n=\H,\V}
	\langle 
		\mu_{\delta \bar\epsilon}^{\lambda_4} 
		\mu_{\gamma \bar\epsilon}^{\lambda_3}
		\mu_{\beta  0}^{\lambda_2}
		\mu_{\alpha 0}^{\lambda_1} 
	\rangle 
\notag \\
    &\quad\times
	G_{\bar\epsilon \delta}[\omega]
	\int^\infty_0 ds_1 e^{i(\omega_\p-\omega)s_1}
	F_{\gamma\delta\gets\alpha\beta}
	^{{\lambda_4\lambda_3\lambda_2\lambda_1}}
	(\omega,\dt;s_1)
	G_{\alpha 0}[\omega_\p - \omega].
	\label{eq:nonrephasing-ESA}
\end{align}
The second terms in Eqs.~\eqref{eq:rephasing-SE2}, \eqref{eq:rephasing-GSB}, \eqref{eq:nonrephasing-SE}, and \eqref{eq:nonrephasing-GSB} originate from the field commutator \cite{Ishizaki:2020jl-SI}. These terms do not depend on $\dt$; therefore, their contributions to the signal can be ignored through the consideration of the difference spectrum.

%%%%%%%%%%%%%%%%%%%%
\section*{S3. Discussion of two-particle resonances}

Here, we argue that the transmission signal in Eq.~(5) in the main text should exhibit no collective two-particle contribution as found in Refs.~\citenum{Muthukrishnan:2004in-SI} and \citenum{Richter:2011co-SI}.
We consider two noninteracting molecules coupled to the light field.
We assume that the time evolution in the $t_1$ and $t_3$ periods is described as $G_{\alpha\beta}(t)=e^{-i\omega_{\alpha\beta}t-\lambda t}$.
The matrix element of time-evolution operator in the $t_2$ period is also modeled as $G_{\alpha\beta\gets\alpha\beta}(t_2)=e^{-i\omega_{\alpha\beta}t_2-\Gamma_{\alpha\beta} t_2}$, where $\Gamma_{\alpha\beta} = 0$ for $\alpha = \beta$ and $\Gamma_{\alpha\beta}=\lambda$ for $\alpha \neq \beta$.
In the limit of $T_\e\to 0$, the expression of $F_{\gamma\delta\gets\alpha\beta}^{{\lambda_4\lambda_3\lambda_2\lambda_1}}(\omega,\dt;0)$ in Eq.~(9) in the main text is obtained as
\begin{align}
	F_{\alpha\beta\gets\alpha\beta}	
	^{{\lambda_4\lambda_3\lambda_2\lambda_1}}(\omega,\dt;0) 
	&\propto
	e^{-i\omega_{\alpha\beta}\dt-\Gamma_{\alpha\beta} \dt} 		
	\delta_{\lambda_1\H}
	\delta_{\lambda_2\H}
	\delta_{\lambda_3\V}
	\delta_{\lambda_4\V}.
	\label{eq:F-func-dt0-dimer}
\end{align}
Inserting Eq.~\eqref{eq:F-func-dt0-dimer} into Eq.~\eqref{eq:rephasing-ESA}, we obtain
\begin{align}
	S_{\rephasing, \ESA}(\omega,\dt ;\omega_\p)
	&=
	\eta
	\Re
	\sum_{\alpha\beta \bar\epsilon}
	\sum_{\lambda_n=\H,\V}
	\langle 
		\mu_{\beta \bar\epsilon}^{\V} 
		\mu_{\alpha \bar\epsilon}^{\lambda_3}
		\mu_{\beta  0}^{\lambda_2}
		\mu_{\alpha 0}^{\lambda_1} 
	\rangle 
	G_{\bar\epsilon \beta}[\omega]
	F_{\alpha\beta\gets\alpha\beta}
	^{{\V\lambda_3\lambda_2\lambda_1}}
	(\omega,\dt;0)
	G_{0 \beta}[\omega_\p - \omega]
\notag \\
	&\propto
	-
	\eta
	\Re
	\sum_{\alpha\beta \bar\epsilon}
	\langle 
		\mu_{\alpha 0}^{\V} 
		\mu_{\beta 0}^{\V}
		\mu_{\beta  0}^{\H}
		\mu_{\alpha 0}^{\H} 
	\rangle 
	\frac{e^{-i\omega_{\alpha\beta}\dt-\Gamma_{\alpha\beta} \dt} }{(\omega-\omega_{\alpha 0}+i\lambda)(\omega_\p-\omega-\omega_{\beta 0}+i\lambda)}.
	\label{eq:no-resonances}
\end{align}
Note $\omega_{\bar\epsilon\beta}=\omega_{\alpha 0}$, $\mu_{\alpha \bar\epsilon}^{\lambda_3}=\mu_{\beta 0}^{\lambda_3}$, and $\mu_{\beta \bar\epsilon}^{\lambda_4}=\mu_{\alpha 0}^{\lambda_4}$ because of the noninteracting dimer system.
Equation~\eqref{eq:no-resonances} represents single-particle resonances, where the two molecules are excited individually.
In other words, the occurrence of simultaneous excitation of two independent molecules by the entangled photons does not occur.
Similarly, the SE and GSB contributions are also understood.
Therefore, the signal in Eq.~(5) exhibits no collective two-particle contribution.
This result is consistent with the arguments in Refs.~\citenum{Muthukrishnan:2004in-SI} and \citenum{Richter:2011co-SI}.

%%%%%%%%%%%%%%%%%%%%
\section*{S4. Birefringent phase-matching}

We consider the PDC process in a birefringent crystal.
Here, we use a negative uniaxial nonlinear crystal such as $\beta$-barium borate (BBO).
The crystal is assumed to have an infinite extent in the $x$ and $y$ directions, and a width $L$ in the $z$ direction.
We also assumed that the pump beam propagates in the $z$ direction.

In birefringent crystals, the refractive index for ordinary polarization is independent of the direction, whereas that of extraordinary polarization depends on the propagation angle of the beam with respect to the optical axis.
It is determined by \cite{Simon:2017qu}
\begin{align}
    \frac{1}{n_{\mathrm{e}}(\omega,\theta)^2}
    =
    \frac{\cos^2\theta}{n_{\mathrm{o}}(\omega)^2}
    +
    \frac{\sin^2\theta}{n_{z}(\omega)^2},
\end{align}
where the refractive indices $n_{\mathrm{o}}(\omega)$ and $n_{z}(\omega)$ are given by the Sellemeier equations for the crystal.
For example, the Sellemeier equations for a BBO crystal \cite{Kato:1986se} are given in the section~S6.

In general, in birefringent phase matching there can be eight possible polarization scenarios for the pump, signal, and idler photons.
For negative uniaxial crystals, the polarization of the pump laser needs to be extraordinary to satisfy the phase-matching condition because $n_{\mathrm{e}}<n_{\mathrm{o}}$ and $n_x=n_y=n_{\mathrm{o}}$.
In type-II PDC ($\mathrm{e} \to \mathrm{o} + \mathrm{e}$), the wave vector mismatch, $\Delta k(\omega_1,\omega_2,\theta)$, is represented by a combination of the following equations:
\begin{align}
    \Delta k(\omega_1,\omega_2;\theta)
    &=
    k_{\mathrm{e}}(\omega_1+\omega_2,\theta)
    -
    k_{\mathrm{o}}(\omega_1)
    \cos \phi_1
\notag \\
    &\quad
    -
    k_{\mathrm{e}}(\omega_2,\theta + \phi_2)
    \cos \phi_2,
	\label{eq:birefringent-phase-matching}
\end{align}
\begin{align}
    k_{\mathrm{o}}(\omega_1)
    \sin \phi_1
    +
    k_{\mathrm{e}}(\omega_2,\theta+\phi_2)
    \sin \phi_2
    =0,
    \label{eq:birefringent-phase-matching2}
\end{align}
where
\begin{align}
    k_{\mathrm{o}}(\omega)
    =
    \frac{\omega n_{\mathrm{o}}(\omega)}{c},
\end{align}
\begin{align}
    k_{\mathrm{e}}(\omega,\theta)
    =
    \frac{\omega n_{\mathrm{e}}(\omega,\theta)}{c},
\end{align}
and $c$ is the speed of light in vacuum.
The angle $\phi_1$ ($\phi_2$) is the scattering angle of the ordinary (extraordinary) beam with respect to the pump beam direction.
In the collinear configuration, where $\phi_1=\phi_2=0$, the wave vector mismatch in Eqs.~\eqref{eq:birefringent-phase-matching} and \eqref{eq:birefringent-phase-matching2} simplifies to
\begin{align}
    \Delta k(\omega_1,\omega_2;\theta)
    &=
    k_{\mathrm{e}}(\omega_1+\omega_2,\theta)
    -
    k_{\mathrm{o}}(\omega_1)
    -    
    k_{\mathrm{e}}(\omega_2,\theta).
\end{align}

%%%%%%%%%%%%%%%%%%%%
\section*{S5. Quasi-phase-matching}

Another phase-matching technique is quasi-phase matching \cite{Boyd:2003no-SI}.
The idea is to achieve phase matching using a multidomain material that periodically reverses the sign of nonlinear susceptibility.
In type-II quasi-phase matching ($\mathrm{e} \to \mathrm{o} + \mathrm{e}$), the wave vector mismatch is expressed as
\begin{align}
    \Delta k(\omega_1,\omega_2;\Lambda)
    &=
    k_{\mathrm{e}}(\omega_1+\omega_2)
    -
    k_{\mathrm{o}}(\omega_1)
    -
    k_{\mathrm{e}}(\omega_2)
    -
    \frac{2\pi}{\Lambda},
\end{align}
where $\Lambda$ is the poling period. 
In contrast to birefringent phase matching, where the phase-matching condition is achieved by tuning the propagation angle of the pump beam with respect to the optical axis, quasi-phase matching works by adjusting the poling period, $\Lambda$.

%%%%%%%%%%%%%%%%%%%%
\section*{S6. Beta-barium borate (BBO)}

For $\beta$-barium borate (uniaxial: $n_x=n_y=n_{\mathrm{o}}$), we used the following Sellmeier equations for the ordinary and extraordinary indices \cite{Kato:1986se-SI}:
\begin{align}
    n_{\mathrm{o}}^2(\lambda)
    =
    2.7359
    +
    \frac{0.01878}{\lambda^2 - 0.01822}
    -
    0.01354 \lambda^2,
\end{align}
\begin{align}
    n_{z}^2(\lambda)
    =
    2.3753
    +
    \frac{0.01224}{\lambda^2 - 0.01667}
    -
    0.01516 \lambda^2.
\end{align}
Here, $\lambda$ is the wavelength of light in micrometers.

%%%%%%%%%%%%%%%%%%%%
\section*{S7. Periodically poled potassium titanyl phosphate (PPKTP)}

For potassium titanyl phosphate, the Sellmeier equations for $n_y$ and $n_z$ are given by \cite{Fradkin:1999tu-SI,Konig:2004ex-SI}
\begin{align}
    n_{y}^2(\lambda)
    =
    2.09930
    +
    \frac{0.922683 \lambda^2}{\lambda^2 - 4.67695 \times 10^{-2}}
    -
    1.38408 \times 10^{-2} \lambda^2,
\end{align}

\begin{align}
    n_{z}^2(\lambda)
    &=
    2.12725
    +
    \frac{1.18431 \lambda^2}{\lambda^2 - 5.14852 \times 10^{-2}}
    +
    \frac{0.6603 \lambda^2}{\lambda^2 - 100.00507}
    -
    9.68956 \times 10^{-3} \lambda^2,
\end{align}
where $\lambda$ is the wavelength of light in micrometers.
The temperature dependence is taken into account by an additional term $\Delta n_j(\lambda,T_{\rm KTP})$ such that
\begin{align}
    n_j(\lambda,T_{\rm KTP})
    =
    n_{j}(\lambda)
    +
    \Delta n_j(\lambda,T_{\rm KTP}).
\end{align}
The temperature-dependent term is given by
\begin{align}
    \Delta n_j(\lambda,T_{\rm KTP})
    =
    n_{j,1}(\lambda) (T_{\rm KTP} - 298)
    +
    n_{j,2}(\lambda) (T_{\rm KTP} - 298)^2,
\end{align}
where the unit of $T_{\rm KTP}$ is K and $n_{j,l}(\lambda)$ is defined as
\begin{align}
    n_{j,l}(\lambda)
    =
    \sum_{m=0}^3 \frac{a_{j,m}}{\lambda^m}.
\end{align}
We used the components $a_{j,m}$ for the $j$ axis ($j=y,z$) given in Ref.~\citenum{Emanueli:2003te}.

The thermal expansion of the crystal and poling period is governed by a parabolic dependence on the temperature \cite{Emanueli:2003te}:
\begin{align}
    L(T_{\rm KTP})
    =
    L_0
    [1+\alpha_{x,0} (T_{\rm KTP} - 298) + \alpha_{x,1} (T_{\rm KTP} - 298)^2 ],
\end{align}
\begin{align}
    \Lambda(T_{\rm KTP})
    =
    \Lambda_0
    [1+\alpha_{x,0} (T_{\rm KTP} - 298) + \alpha_{x,1} (T_{\rm KTP} - 298)^2 ]
\end{align}
with $\alpha_{x,0}=6.7 \times 10^{-6}$ and $\alpha_{x,1}=11.0 \times 10^{-9}$.

%%%%%%%%%%%%%%%%%%%%
\section*{S8. Case of squeezed light}

The squeezed vacuum state of the field is given by
\begin{align}
	\lvert \psi_\twin \rangle
	=
	\hat{U}
	\vert \vac \rangle,
\end{align}
where
\begin{align}
	\hat{U}
	=
	\exp
	\left(
	\int d\omega
	r(\omega)
	\adagger_\H(\omega) \adagger_\V(\omega_\p - \omega)
	\vert \vac \rangle
	-
	{\rm h.c.}
	\right),
\end{align}
\begin{align}
	r(\omega)
	=
	\zeta
	\Phi(\omega_\p -\omega - \bar\omega_\V).
\end{align}
In the Heisenberg picture, the operators $\a_\H(\omega)$ and $\a_\V(\omega)$ transform as
\begin{align}
	\a_\H(\omega)
	&\to
	\hat{U}^\dagger
	\a_\H(\omega)
	\hat{U}
\notag \\
	&=
	\cosh r(\omega)
	\a_\H(\omega)
	+
	\sinh r(\omega)
	\adagger_\V(\omega_\p - \omega),
\end{align}
\begin{align}
	\a_\V(\omega)
	&\to
	\hat{U}^\dagger
	\a_\V(\omega)
	\hat{U}
\notag \\
	&=
	\cosh r(\omega)
	\a_\H(\omega)
	+
	\sinh r(\omega)
	\adagger_\H(\omega_\p - \omega).
\end{align}
When the entanglement time is sufficiently short compared to characteristic timescales of the dynamics under investigation, the signal with the squeezed vacuum state is expressed as
%Thus, the signal with the squeezed vacuum state is expressed as
\begin{align}
	S(\omega,\dt ;\omega_\p)
	&=
	S_{\rephasing, \SE}(\omega,\dt ;\omega_\p)
	+
	S_{\nonrephasing, \SE}(\omega,\dt ;\omega_\p)
\notag \\
	&\quad 
	+
	S_{\rephasing, \GSB}(\omega,\dt ;\omega_\p)
	+
	S_{\nonrephasing, \GSB}(\omega,\dt ;\omega_\p)
\notag \\
	&\quad 
	+
	S_{\rephasing, \ESA}(\omega,\dt ;\omega_\p)
	+
	S_{\nonrephasing, \ESA}(\omega,\dt ;\omega_\p),
	\label{eq:signal-squeezed}
\end{align}
where
\begin{align}
	S_{\rephasing, \SE}(\omega,\dt ;\omega_\p)
	&=	
	-
	\Re
	\frac{1}{\hbar^3}
	\calE(\bar\omega_\H)^2 \calE(\bar\omega_\V)^2
	\sum_{\alpha\beta\gamma\delta}
	\sum_{\lambda_n=\H,\V}
	\langle 
		\mu_{\delta 0}^{\lambda_4} 
		\mu_{\gamma 0}^{\lambda_3}
		\mu_{\beta  0}^{\lambda_2}
		\mu_{\alpha 0}^{\lambda_1} 
	\rangle 
%	e^{-i\omega t}
	G_{\gamma 0}[\omega]
\notag \\
	&\quad \times
	\left[
	\sinh^4\zeta
	\,
	\delta_{\lambda_1\H}
	\delta_{\lambda_2\V}
	\delta_{\lambda_3\H}
	\delta_{\lambda_4\V}
	\right.
\notag \\
	&\quad
	+
	G_{\gamma\delta \leftarrow \alpha\beta}[\omega=0]
%	G_{0\alpha}(0)
	\sinh^2\zeta
	\cosh^2\zeta
	\,
	\delta_{\lambda_1\H}
	\delta_{\lambda_2\H}
	\delta_{\lambda_3\V}
	\delta_{\lambda_4\V}
\notag \\
	&\quad
	\left.
	+
	G_{\gamma\delta \leftarrow \alpha\beta}(\dt)
	G_{0\beta} [\omega_\p -\omega]
	\sinh^2\zeta
	\cosh^2\zeta
	\,
	\delta_{\lambda_1\H}
	\delta_{\lambda_2\H}
	\delta_{\lambda_3\V}
	\delta_{\lambda_4\V}
	\right],
	\label{eq:rephasing-SE-squeezed}
\end{align}
\begin{align}
	S_{\nonrephasing, \SE}(\omega,\dt ;\omega_\p)
	&=
	-
	\Re
	\frac{1}{\hbar^3}
	\calE(\bar\omega_\H)^2 \calE(\bar\omega_\V)^2
	\sum_{\alpha\beta\gamma\delta}
	\sum_{\lambda_n=\H,\V}
	\langle 
		\mu_{\delta 0}^{\lambda_4} 
		\mu_{\gamma 0}^{\lambda_3}
		\mu_{\beta  0}^{\lambda_2}
		\mu_{\alpha 0}^{\lambda_1} 
	\rangle 
	G_{\gamma 0}[\omega]
\notag \\
	&\quad \times
	\left[
	G_{\alpha 0}[\omega]
	\sinh^4\zeta	
	\,
	\delta_{\lambda_1\V}
	\delta_{\lambda_2\H}
	\delta_{\lambda_3\H}
	\delta_{\lambda_4\V}
	\right.
\notag \\	
	&\quad 
	+
	G_{\gamma\delta \leftarrow \alpha\beta}[\omega=0]
	\sinh^2\zeta
	\cosh^2\zeta
	\,
	\delta_{\lambda_1\H}
	\delta_{\lambda_2\H}
	\delta_{\lambda_3\V}
	\delta_{\lambda_4\V}
\notag \\	
	&\quad 
	\left.
	+
	G_{\gamma\delta \leftarrow \alpha\beta}(\dt)
	G_{\alpha 0}[\omega_\p -\omega]
	\sinh^2\zeta
	\cosh^2\zeta	
	\,
	\delta_{\lambda_1\H}
	\delta_{\lambda_2\H}
	\delta_{\lambda_3\V}
	\delta_{\lambda_4\V}
	\right],
\end{align}
\begin{align}
	S_{\rephasing, \GSB}(\omega,\dt ;\omega_\p)
	&=	
	-
	\Re
	\frac{1}{\hbar^3}
	\calE(\bar\omega_\H)^2 \calE(\bar\omega_\V)^2
	\sum_{\alpha\beta}
	\sum_{\lambda_n=\H,\V}
	\langle 
		\mu_{\beta 0}^{\lambda_4} 
		\mu_{\beta 0}^{\lambda_3}
		\mu_{\alpha  0}^{\lambda_2}
		\mu_{\alpha 0}^{\lambda_1} 
	\rangle 
%	e^{-i\omega t}
	G_{\beta 0}[\omega]
\notag \\	
	&\quad \times
	\left[
	G_{00 \leftarrow 00}[\omega=0]
	\sinh^4\zeta
	\,
	\delta_{\lambda_1\H}
	\delta_{\lambda_2\H}
	\delta_{\lambda_3\V}
	\delta_{\lambda_4\V}
	\right.
\notag \\	
	&\quad
	+
	G_{00 \leftarrow 00}(\dt)
	G_{0 \alpha} [\omega_\p - \omega]
	\sinh^2\zeta
	\cosh^2\zeta
	\,
	\delta_{\lambda_1\H}
	\delta_{\lambda_2\H}
	\delta_{\lambda_3\V}
	\delta_{\lambda_4\V}
\notag \\	
	&\quad
	\left.
	+
	G_{0 \alpha} [\omega]
	\sinh^2\zeta
	\cosh^2\zeta
	\,
	\delta_{\lambda_1\H}
	\delta_{\lambda_2\V}
	\delta_{\lambda_3\H}
	\delta_{\lambda_4\V}
	\right],
\end{align}
\begin{align}
	S_{\nonrephasing, \GSB}(\omega,\dt ;\omega_\p)
	&=	
	-
	\Re
	\frac{1}{\hbar^3}
	\calE(\bar\omega_\H)^2 \calE(\bar\omega_\V)^2
	\sum_{\alpha\beta}
	\sum_{\lambda_n=\H,\V}
	\langle 
		\mu_{\beta 0}^{\lambda_4} 
		\mu_{\beta 0}^{\lambda_3}
		\mu_{\alpha  0}^{\lambda_2}
		\mu_{\alpha 0}^{\lambda_1} 
	\rangle 
	G_{\beta 0}[\omega]
\notag \\	
	&\quad \times
	\left[	
	G_{00 \leftarrow 00}[\omega=0]
	\sinh^4\zeta
	\,
	\delta_{\lambda_1\H}
	\delta_{\lambda_2\H}
	\delta_{\lambda_3\V}
	\delta_{\lambda_4\V}
	\right.
\notag \\	
	&\quad 
	+
	G_{00 \leftarrow 00}(\dt)
	G_{\alpha 0 }[\omega_\p -\omega]
	\sinh^2\zeta
	\cosh^2\zeta
	\,
	\delta_{\lambda_1\H}
	\delta_{\lambda_2\H}
	\delta_{\lambda_3\V}
	\delta_{\lambda_4\V}
\notag \\	
	&\quad 
	\left.
	+
	G_{\alpha 0 }[\omega]
	\sinh^2\zeta
	\cosh^2\zeta
	\,
	\delta_{\lambda_1\V}
	\delta_{\lambda_2\H}
	\delta_{\lambda_3\H}
	\delta_{\lambda_4\V}
	\right],
\end{align}	
\begin{align}
	S_{\rephasing, \ESA}(\omega,\dt ;\omega_\p)
	&=
	\Re
	\frac{1}{\hbar^3}
	\calE(\bar\omega_\H)^2 \calE(\bar\omega_\V)^2
	\sum_{\alpha\beta\gamma\delta \bar\epsilon}
	\sum_{\lambda_n=\H,\V}
	\langle 
		\mu_{\delta \bar\epsilon}^{\lambda_4} 
		\mu_{\gamma \bar\epsilon}^{\lambda_3}
		\mu_{\beta  0}^{\lambda_2}
		\mu_{\alpha 0}^{\lambda_1} 
	\rangle 
	G_{\bar\epsilon \delta}[\omega]
\notag \\	
	&\quad 
	\times
	\left[
	\sinh^4\zeta
	\,
	\delta_{\lambda_1\H}
	\delta_{\lambda_2\V}
	\delta_{\lambda_3\H}
	\delta_{\lambda_4\V}
	\right.
\notag \\	
	&\quad 
	+
	G_{\gamma\delta \leftarrow \alpha\beta}[\omega=0]
	\sinh^4\zeta
	\,
	\delta_{\lambda_1\H}
	\delta_{\lambda_2\H}
	\delta_{\lambda_3\V}
	\delta_{\lambda_4\V}
\notag \\	
	&\quad 	
	\left.
	+
	G_{\gamma\delta \leftarrow \alpha\beta}(\dt)
	G_{0\beta} [\omega_\p -\omega]
	\sinh^2\zeta
	\cosh^2\zeta	
	\,
	\delta_{\lambda_1\H}
	\delta_{\lambda_2\H}
	\delta_{\lambda_3\V}
	\delta_{\lambda_4\V}
	\right],
\end{align}
\begin{align}
	S_{\nonrephasing, \ESA}(\omega,\dt ;\omega_\p)
	&=
	\Re
	\frac{1}{\hbar^3}
	\calE(\bar\omega_\H)^2 \calE(\bar\omega_\V)^2
	\sum_{\alpha\beta\gamma\delta \bar\epsilon}
	\sum_{\lambda_n=\H,\V}
	\langle 
		\mu_{\delta \bar\epsilon}^{\lambda_4} 
		\mu_{\gamma \bar\epsilon}^{\lambda_3}
		\mu_{\beta  0}^{\lambda_2}
		\mu_{\alpha 0}^{\lambda_1} 
	\rangle 
	G_{\bar\epsilon \delta}[\omega]
\notag \\
	&\quad
	\times
	\left[
	G_{\alpha 0}[\omega]	
	\sinh^4\zeta	
	\,
	\delta_{\lambda_1\H}
	\delta_{\lambda_2\V}
	\delta_{\lambda_3\H}
	\delta_{\lambda_4\V}
	\right.	
\notag \\
	&\quad
	+
	G_{\gamma\delta \leftarrow \alpha\beta}[\omega=0]
	\sinh^4\zeta
	\,
	\delta_{\lambda_1\H}
	\delta_{\lambda_2\H}
	\delta_{\lambda_3\V}
	\delta_{\lambda_4\V}
\notag \\
	&\quad
	\left.
	+
	G_{\gamma\delta \leftarrow \alpha\beta}(\dt)
	G_{\alpha 0}[\omega_\p -\omega]
	\sinh^2\zeta
	\cosh^2\zeta
	\,
	\delta_{\lambda_1\H}
	\delta_{\lambda_2\H}
	\delta_{\lambda_3\V}
	\delta_{\lambda_4\V}
	\right].
	\label{eq:nonrephasing-ESA-squeezed}
\end{align}
The first term in Eq.~\eqref{eq:rephasing-SE-squeezed} is the incoherent contribution induced by the squeezed vacuum state, which has no temporal resolution.
Since this term is independent of the frequency of the delay time, $\dt$, this contribution to the total signal in Eq.~\eqref{eq:signal-squeezed} can be removed by considering the difference spectrum $\Delta S(\omega,\dt ;\omega_\p)$ in Eq.~(10). 
Thus, it is demonstrated that it possible to extract relevant information on the excited-state dynamics from the signal even in the case of high-gain squeezed vacuum states ($\zeta \gg 1$).

\clearpage

%%%%%%%%%%%
\begin{figure}
\centering
    \includegraphics{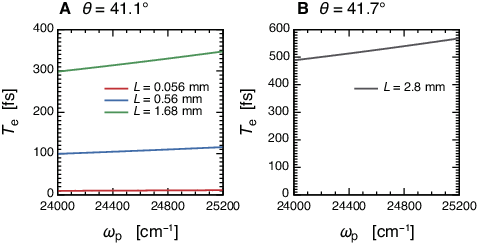}
    \caption{
	The entanglement time as a function $\omega_\p$ of the BBO crystal for (a) $\theta=41.1^\circ$ and (b) $\theta=41.7^\circ$.
	The scattering angles are set to $\phi_1=\phi_2=0$.}
\end{figure}
%%%%%%%%%%%

\clearpage

%%%%%%%%%%%
\begin{figure}
\centering
    \includegraphics{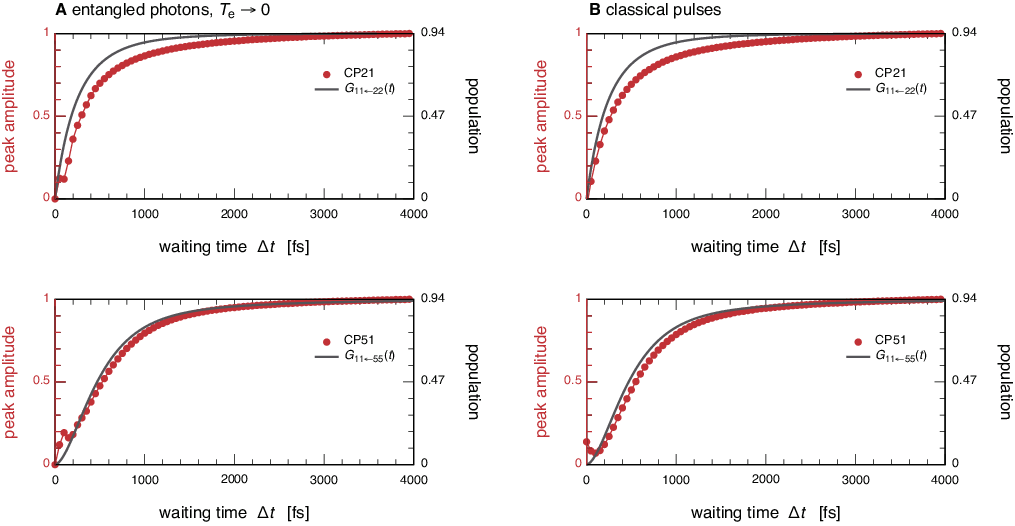}
    \caption{
	(A) Time evolution of the amplitudes of CP21 and CP51 in the difference spectra of the FMO complex with entangled photon pairs in the limit of short entanglement time. The temperature is set to $T=77\,{\rm K}$.
	For comparison, the black line represents the matrix element of the time-evolution operator, $G_{11\gets 22}(t)$ ($G_{11\gets 55}(t)$), calculated directly in the modified Redfield theory, which corresponds to $e_2 \to e_1$ ($e_5 \to e_1$) transport.
	The normalization of the amplitude of CP21 (CP51) is such that the maximum value of the peak amplitude is unity.
	(B) Time evolution of the amplitudes of CP21 and CP51 in absorptive 2D spectra of the FMO complex obtained with the Fourier-transformed photon-echo measurement in the impulsive limit.
	We chose the HHVV sequence as the polarization sequence of the four laser pulses.}
\end{figure}
%%%%%%%%%%%

\clearpage

%%%%%%%%%%%
\begin{figure}
\centering
    \includegraphics{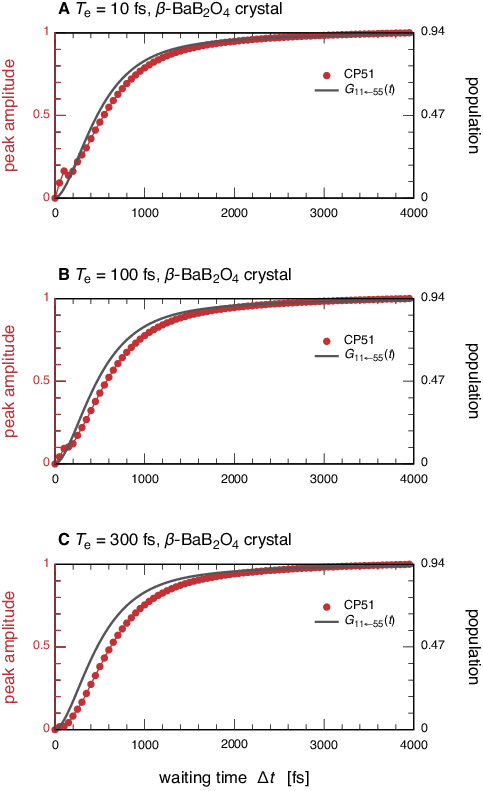}
    \caption{
	Time evolution of the amplitudes of CP51 in the difference spectra of the FMO complex with entangled photon pairs generated via the BBO crystal for (A) $T_\e = 10\,{\rm fs}$ (i.e., $L = 0.056\,{\rm mm}$), (B) $T_\e = 100\,{\rm fs}$ ($L = 0.56\,{\rm mm}$), and (C) $T_\e = 300\,{\rm fs}$ ($L = 1.68\,{\rm mm}$). 
	The propagation angle of the beam with respect to the optic axis is set to $\theta=41.1^\circ$.
	The scattering angles are set to $\phi_1=\phi_2=0$.
	The other parameters are the same as that in Fig.~4 in the main text.
	For comparison, the black line represents the matrix element of the time-evolution operator, $G_{11\gets 55}(t)$, calculated directly in the modified Redfield theory, which corresponds to $e_5 \to e_1$ transport. The normalization of the amplitude of CP51 is such that the maximum value of the peak amplitude is unity.
    }
\end{figure}
%%%%%%%%%%%

\clearpage

%%%%%%%%%%%
%%%%%%%%%%%

\end{document}